\documentclass[aps,prb,twocolumn,groupedaddress,amsmath,amssymb,showpacs,
eqsecnum,floatfix]{revtex4}

\usepackage{graphicx}
\begin{document}


\title{Melting and structure of the vortex solid in strongly anisotropic 
layered  superconductors with  random columnar pins}

\author{Chandan Dasgupta}
\email[]{cdgupta@physics.iisc.ernet.in}
\affiliation{Department of Physics, Indian Institute of Science, 
Bangalore 560012, India}
\author {Oriol T. Valls}
\email[]{otvalls@umn.edu}
\affiliation{School of Physics and Astronomy and Minnesota Supercomputer
Institute, University of Minnesota, Minneapolis, Minnesota 55455}


\date{\today}

\begin{abstract}
We study  the melting transition 
of the low-temperature vortex solid in strongly
anisotropic layered superconductors with a  concentration of random
columnar pinning centers small enough so that
the areal density of the pins is  much less than that of the
vortex lines. Both the external magnetic field and the 
columnar pins are assumed to be  oriented perpendicular to the layers
Our method, involving numerical minimization of
a model free energy functional, yields not only the
free energy values at the local minima of the functional
but also the detailed density
distribution of the system at each minimum: this allows us
to study in detail the structure
of the different phases. 
We find that at these pin concentrations
and low temperatures, the thermodynamically stable state
is a topologically ordered Bragg glass.
This nearly crystalline state melts into an interstitial liquid (a liquid
in which a small fraction of vortex
lines remain localized at the pinning centers)
in two steps, so that the Bragg glass and the liquid are separated by a
narrow phase that we identify from analysis of its density
structure as a polycrystalline Bose glass. Both the Bragg glass to Bose glass
and the Bose glass to interstitial liquid transitions are first-order.
We also find that a local melting temperature defined using a criterion based
on the degree of localization of the vortex lines exhibits
spatial variations similar to those observed in recent experiments. 
\end{abstract}


\maketitle

\section{Introduction\label{intro}}

The mixed phase of type-II superconductors 
with random pinning is generally regarded to be an
archetypal test system for the study of the effects of quenched disorder
on the structure and melting of solids. In this phase,
magnetic flux penetrates
the sample as quantized vortex lines which form a special physical
system known as ``vortex matter''. The fascinating equilibrium and
dynamical properties of vortex matter in
the mixed phase of high-temperature superconductors (HTSCs)
have prompted considerable experimental and theoretical
attention for more than a decade (see Ref.~\onlinecite{blat}
for an early review). 
The  mixed phase of HTSCs is
strongly affected by the presence of
pinning centers, either intrinsic or artificially
generated.  Understanding the effects of pinning in these systems is
basic for practical applications because 
pinning strongly influences the value of the critical current
\cite{dg}.

The vortex lines in a pure type-II superconductor
form a triangular lattice (the
Abrikosov lattice) at low temperatures. Because of 
enhanced thermal fluctuations in 
highly anisotropic, layered HTSCs, the Abrikosov lattice in very pure
samples, in a magnetic field perpendicular to the 
layers (the field will be assumed to be in 
this direction throughout our discussion), undergoes
a first-order melting transition\cite{blat} into a resistive 
vortex liquid (VL) as the temperature $T$
increases. Random pinning destroys the long-range
translational order of the Abrikosov lattice 
and leads to the occurrence of a variety of glassy phases at low $T$.
It is now generally accepted that
in systems with random point pinning, a topologically
ordered low-temperature phase with quasi-long-range translational order 
(denoted  as the ``Bragg glass'' (BrG) phase) occurs 
at low fields if the pinning disorder is sufficiently weak. This has been 
established theoretically
\cite{nat,gia,ns}
as well as experimentally (see e.g. Ref.~\onlinecite{kle}).
The possibility  of an amorphous 
vortex glass (VG) phase, with nonlinear voltage-current
characteristics and vanishing resistance in the zero-current limit, in
systems with strong pinning (or at high magnetic fields where the effects
of pinning disorder are enhanced) was suggested by Fisher\cite{fis} (see also
Ref.~\onlinecite{ffh}). However, in spite of extensive
investigations, 
the existence of a true VG phase (i.e., an amorphous glassy phase
thermodynamically distinct
from the high-temperature VL) in systems with uncorrelated
point pinning remains very controversial:  
different calculations
\cite{bok,ves}
lead to different conclusions and the experimental situation\cite{pet,str}
is similarly contradictory.
A variety of  ``glassy'' behavior has been reported in different
experiments 
(see Ref.~\onlinecite{ban1} and references therein) near the first-order 
melting transition of the BrG phase of both conventional superconductors and
HTSCs with point pinning. 
It has been suggested\cite{ban1,men}
that these observations may be understood if it is assumed that the
melting of the BrG phase occurs in two steps: the BrG first transforms into
a ``multidomain'' glassy phase that melts into the usual VL at a
slightly higher $T$.

Columnar pinning defects can be produced
for example as damage tracks arising from
heavy-ion bombardment.  The technological importance of these
extended defects oriented parallel to
the direction of the external magnetic field has been long 
recognized\cite{civ,bud}:
they are  highly effective in increasing the
critical current by localizing vortex lines along their length.
Heavy-ion irradiation  produces a random array of parallel
columnar defects each of which can trap one or more vortices at low
temperatures. The effects of such random arrays of extended defects, 
oriented perpendicular to the superconducting layers, on
the properties of the mixed phase of HTSCs have been extensively
studied  experimentally\cite{kha,ban,meng}. The same question has also
been examined theoretically\cite{nel,rad,lar,vin}
and numerically\cite{wen,sen,tya,non}.
When the columnar pinning is strong, and their concentration exceeds that
of the vortex lines, a 
so-called ``strong'' Bose Glass (BoG) phase with nearly all the vortex lines
localized at the pinning centers occurs\cite{nel,wen} at low $T$. 
This phase is strongly disordered with very short-range translational and
bond-orientational correlations.
The behavior in this regime is
fairly well-understood in terms of a mapping\cite{nel} of the
thermodynamics of the system to the quantum mechanical properties of a 
two-dimensional system of interacting bosons in a random potential.

Much less is known about the behavior in the dilute pin limit where 
the concentration of pins is
much smaller than that of vortex lines. In such systems, 
one expects\cite{rad} a ``weak'' BoG phase at low         
temperatures, with a small fraction of vortex lines localized strongly at
the pinning centers and the remaining ones localized relatively weakly in
the interstitial region between pinning centers. As the temperature is
increased, this phase should\cite{rad,vin} melt
into an interstitial liquid (IL). In the IL         
phase, some of the vortices remain trapped at the pinning centers,         
while the other, interstitial ones, form a liquid. The pinned vortices 
are expected\cite{rad,vin} 
to delocalize, thereby forming the usual VL, at a crossover
occuring at a higher temperature. The melting transition 
of the ``weak'' BoG phase into
the IL is predicted\cite{rad,lar} to be first-order for small 
pin concentrations, whereas the ``strong'' BoG to VL transition occurring for
large pin concentrations is known\cite{nel,lar} to be a continuous one.  

A first-order melting transition of the ``weak'' BoG phase
has been observed\cite{meng} in recent experiments on samples of
${\rm Bi_2Sr_2CaCu_2O_{8+x}}$ (BSCCO) with a small concentration of columnar
pins. The BoG phase in these systems is found\cite{ban,meng} to have a 
polycrystalline structure with ordered vortex crystallites of different
orientations embedded in the interstitial region between vortices pinned at
the columnar defects. If the pin concentration is sufficiently small, the
melting of this BoG phase upon increasing the temperature occurs into an
IL phase with a fraction of vortex lines remaining pinned at the defects. 
These pinned vortices delocalize at a higher temperature\cite{ban}. The
temperature at which the first-order BoG to IL transition occurs  
is\cite{kha,ban} very close to the melting temperature of the
same system without the columnar pins, if the pin concentration is small.
As the pin concentration is increased, the melting temperature increases and 
the transition eventually becomes continuous. The difference
between the melting temperature and the temperature at which the pinned 
vortices in the IL phase delocalizes decreases and eventually goes to zero
as the pin concentration increases. Another interesting feature, found
experimentally for both point\cite{soi}
and columnar\cite{ban}         
pinning, is that the melting of the low $T$ ``solid'' phase is ``inhomogeneous''
in that it occurs locally over a range of temperatures: 
the local transition temperature, which can be
measured from the local magnetization, is different in different
regions of the sample. This inhomogeneity of the local melting temperature is
believed\cite{soi} to be closely related to the local arrangement of the
specific pinning centers (the so-called ``pinning landscape'') in each
sample studied. 

The polycrystalline nature of the BoG phase for small pin concentrations has
also been observed in simulations\cite{sen}.  
Recent numerical work\cite{non} indicates that a BrG phase, with topological
order, may also exist 
in such systems provided that the pin concentration is         
sufficiently small. In these simulations, the BrG phase is found to melt into
the IL phase via a first-order transition as the temperature is increased.
As the pin concentration is increased beyond a 
critical value, the BrG phase disappears\cite{non} and a low-temperature BoG
phase with a continuous BoG to IL transition upon increasing 
the temperature is found\cite{tya,non}. 

In this paper, we approach these problems through a different numerical
method. We consider a layered, strongly anisotropic superconductor
(such as
BSCCO) with a dilute random array of columnar defects oriented perpendicular
to the layers, and with a magnetic
field applied parallel to the columnar pins. We describe the equilibrium 
properties of this system in terms of vortex density variables, using
a free energy functional of the Ramakrishnan-Yussouff\cite{ry} (RY) form.
We numericall minimize a spatially discretized version of this free 
energy functional 
and obtain the vortex density 
configuration at each local minimum. 
Analysis of these density configurations, both in terms
of a variety of correlation functions and by direct
visualization of the arrangement of the vortices (vortex positions are 
identified with those of local peaks of the vortex density),
allows us to identify the nature of the phases corresponding to different
local minima of the free energy.
Comparison of  the values of the free energy at the 
minima as the temperature varies yields the transition
temperatures. 
A similar procedure has been successfully employed for
the case of a regular array\cite{dv1,dv2} of columnar pinning centers.

We find in our study that that a Bragg glass phase exists at low temperatures
for samples with a small concentration of columnar pins. We will present
evidence for this phase from our analysis of the vortex density and associated
correlation functions. Upon warming, this phase melts
into what we show to be an insterstitial liquid (IL) phase,
but the melting   
is shown\cite{dv3}
to occur in two steps: the BrG and IL phases are separated, over a narrow
temperature range, by an intermediate phase which exhibits a
multi-domain polycrystalline structure. We show that this intermediate phase 
should be identified
as a Bose glass (BoG). The temperature of the upper 
(BoG to IL) transition is approximately independent
of the columnar pin concentration $c$, at the low $c$ values studied. 
Both the BrG to BoG and the BoG to IL transitions are found to be 
first-order. We also
find that a local melting temperature can be suitably defined using a 
criterion based on the degree of localization of the vortices, and that
its behavior (in particular, its spatial variation) 
is quite consistent with what is seen in experiments\cite{ban,soi}. The value
of the local melting temperature is strongly correlated with the
presence of topological defects (dislocations) in the vortex solid which, in
turn, is correlated with the local arrangement of the pinning centers. We 
also show that the transition to the IL phase corresponds to a percolation
of regions containing liquid-like (delocalized) vortices across the sample. 

After this Introduction, we discuss our definitions, model and numerical 
methods in the next Section. Then in Section \ref{results} we present
our results. We discuss first the free energy minima and their study
through the use of tools such
as correlation functions, local peak-density plots, and
Voronoi construction for the lattice formed by local density peaks.
These tools
allow the identification of the phases at each free
energy minimum, as we shall show.
Then we derive the phase
diagram and show that indeed the BrG and IL phases are separated  by a thin
sliver of BoG. Finally, our results for 
the nature of the local melting are displayed and discussed.
A brief conclusions Section recapitulates the main points
of the paper and discusses them in the context of existing  results.

\section{Methods\label{methods}}

The general procedures that we use are quite similar to those employed
in previous work on a regular array of columnar pins\cite{dv1,dv2}. We
will therefore give here only a brief summary, emphasizing the details
that are different in the random case considered here.

The system we study is a layered superconductor in the extreme anisotropic
limit, that is, with vanishing Josephson coupling between layers, which
are then coupled via the electromagnetic interaction only. This 
limit is appropriate for the Bi- and Tl- based HTSC compounds in a large
region of the magnetic field ($H$)--$T$ plane. 
In this work, we will use material parameter values
appropriate to BSCCO. With these
assumptions, one can write the energy of a system of ``pancake'' vortices
residing in the layers as a sum of anisotropic two-body
interactions. For straight columnar pins normal to the layers, the pinning 
potential is the same in every layer. It is then possible to write the
free energy as a functional of the {\it time averaged} areal vortex density
$\rho({\bf r})$:
\begin{equation}
F[\rho]-F_0=F_{RY}[\rho]+F_p[\rho] .
\label{fe}
\end{equation}
where $F_0$ is the free energy corresponding to a uniform
vortex liquid of density $\rho_0=B/\Phi_0$ ($B$ is the magnetic induction and
$\Phi_0$ the superconducting flux quantum).
The first term in the right-hand
side of Eq.~(\ref{fe}) is the free energy functional
in the absence of pinning. As explained above we use for this free energy the 
RY\cite{ry} functional, which is known\cite{dv1,dv2,seng,men1} to give
a quantitatively accurate description of the vortex-lattice melting
transition in the absence of pinning. It is of the form:
\begin{eqnarray}
\beta F_{RY}[\rho] &=& \int d^2r\left[\rho({\bf r})
\{ \ln(\rho({\bf r}))-\ln (\rho_0)\} -
\delta\rho({\bf r})\right] \nonumber \\ 
&-&\frac{1}{2}\int d^2r \int d^2r^\prime \,\tilde{C}(|{\bf r} -
{\bf r}^\prime |)
\delta\rho({\bf r}) \delta\rho({\bf r}^\prime). 
\label{ryfe}
\end{eqnarray}
Here $\beta$ is the inverse temperature, 
$\delta\rho({\bf r}) \equiv \rho({\bf r}) - \rho_0$,
and $\tilde{C}(r)$ is the usual direct
pair correlation function\cite{han} which may
be written as a sum over layers: $ \tilde{C}(r)\equiv \sum_n C(n,r)$,
with $C(n,r)$ (where $n$ is the layer
separation and $r$ the in-layer distance) being the corresponding
direct pair correlation function of a
layered liquid of pancake vortices. The direct correlation function,
which is needed as input in our free energy, can be
accurately calculated in a number of ways. We will use here the
results of the hypernetted chain calculation of Ref.~\onlinecite{men1}.
In general, the results in the limit considered depend on the values of the
in-plane London penetration length $\lambda(T)$, the interlayer spacing $d$
and a 
dimensionless coupling parameter $\Gamma$ given by:
\begin{equation}
\Gamma = \beta d \Phi^2_0/8 \pi^2 \lambda^2(T).
\label{gamma}
\end{equation}
For BSCCO we will take $d=15 \AA$, and
we will assume a standard two-fluid temperature dependence for $\lambda(T)$
with $\lambda(0)= 1500 \AA$ and $T_c=85$K (at zero field).

The second term in the right-hand side of Eq.~(\ref{fe}) represents the pinning
and is of the form:
\begin{equation}
F_p[\rho] = \int d^2r V_p({\bf r}) \delta\rho({\bf r}). 
\label{pin}
\end{equation}
where $V_p$ is the pinning potential which can be written as
\begin{equation}
V_p({\bf r})=\sum_j V_0(|{\bf r}-{\bf R}_j|),
\label {pinpot}
\end{equation}
with the sum extending over the planar positions of the random pinning centers.
We take the potential $V_0$ corresponding to a single pinning center
to be of the usual truncated parabolic form:\cite{daf}
\begin{equation}
\beta V_0(r)=-\alpha\Gamma[1-(r/r_0)^2]\Theta(r_0-r)
\label{single}
\end{equation}
where $r_0$ is the range. The basic length in the problem, which
we will use as our unit of length unless otherwise indicated,
is $a_0$, defined by the relation $\pi a_0^2 \rho_0=1$.
We choose $r_0$ to be $r_0=0.1a_0$ and take the dimensionless
constant $\alpha=0.05$, for which value, as previously shown,\cite{dv2}
each pinning center traps approximately one vortex in BSCCO, in the
temperature range of interest. This range is determined by the
following considerations. We will keep the field fixed at $B=0.2$T, and vary
the temperature. The melting temperature of the unpinned lattice is then 
$T_m^{0} \simeq 18.4$K\cite{dv2}, and we, therefore, consider the neighborhood
of this temperature. Then one has, to a very good approximation
$\Gamma=2650/T$ (with the temperature in Kelvin), while 
$\lambda \approx \lambda(0)$. 

To carry out our numerical calculations, we discretize the density variables
on a computational triangular grid of lattice spacing $h$, containing $N^2$
sites. We take the spacing $h$  to be  $h=a/16$ in our calculations,
where $a=1.998 a_0$ is the equilibrium lattice constant of the vortex lattice
in the temperature range considered at the indicated
field. This value, as pointed out in
Ref.~\onlinecite{dv2}, is slightly higher than that of a triangular vortex
lattice of density $\rho_0$, which is $(2 \pi/\sqrt{3})^{1/2}a_0$. We define
at each site $j$ on this computational lattice a variable $\rho_j$,
with $\rho_j \equiv \rho({\bf r}_j) v$, where 
$v$ is the area of each computational
unit cell. Results reported here are for $N=1024$, which,
for the chosen value of $h$, corresponds to including
$N_v=$ 4096 vortices in the calculation. Preliminary and
confirmatory results at $N=512$ (1024 vortices)
were also obtained. A number of pinning sites
($N_p$) are put, at random, on some of the computational lattice sites. The
results presented here, which correspond
to the dilute limit, will be primarily for a pin concentration
$c$ of $c=1/64$, that is, a number of pins $N_p=64$ at $N=1024$ (or
$N_p=16$ at $N=512$). Some preliminary results for
$c=1/32$ ($N_p=128$  at $N=1024$,
or $N_p=32$ at $N=512$) will also be discussed briefly. 
With random pins, the results depend on the particular random
pin configuration and
averaging over different such configurations is required. The dependence is
however not strong: as we shall see, averaging over five to ten configurations
is enough to make statistical errors sufficiently small for
our purposes.

To perform our studies, we 
numerically minimize the discretized free energy with respect
to the $N^2$ discretized density variables $\{\rho_i\}$. 
To do so, the interaction
term in the
right side of Eq.~(\ref{ryfe}) must be repeatedly evaluated, 
and since this term
is of a convolution form, 
this is most efficiently done in momentum space, through the use
of efficient Fast Fourier Transform (FFT) routines. This avoids having
to evaluate this term as a double sum
which would be computationally much more cumbersome than performing
the direct and inverse FFT's.
In performing the minimization, one must keep in mind that the variables
$\{\rho_i\}$ must be nonnegative. This precludes the use of
many efficient minimization algorithms. We use a procedure\cite{num} that
ensures that this constraint is satisfied. Numerical minimization
is performed starting with an appropriate initial condition for the
density variables. As explained in Sec.~\ref{results} below, different
local minima may be found, at the same field and temperature, 
depending on the initial
conditions. These minima correspond to different phases.
The minimization
procedure yields not only the value of the free energy at each minimum but
also the detailed vortex density configuration at the minimum found, i.e.
the values of the set $\{\rho_i\}$ at the minimum. It is then 
straightforward to analyze the actual density configuration in several
ways. It is also possible 
to evaluate any desired density correlations. The nature of the phase
corresponding to each local minimum of the free energy can be inferred from
such analysis. The (mean-field) phase diagram is then obtained from a
comparison of the free energies of the different minima as the temperature
$T$ is varied.
The results of carrying out this program are discussed below.

\section{Results\label{results}}

\subsection{Free energy minima}
\label{minima}

As explained above, minima of the free energy are found by starting the
minimization process with appropriate initial conditions. We use three
different kinds of initial conditions. The first kind is 
a uniform density ($\rho_i = \rho_0 v$ for all $i$), 
corresponding to a completely
disordered liquid state. This is typically used to obtain, as we shall see, 
liquid-like states at temperatures near the equilibrium melting temperature
of the pure vortex lattice, 
which is\cite{dv1,dv2} $T_m^0 \simeq 18.4$K for the value of $B$ considered
here. 
The second
kind corresponds to a  crystalline initial state, with values of
$\{\rho_i\}$ as  obtained in Ref.~\onlinecite{dv2} for the pure
vortex lattice.
In our system with a spatially varying
random pinning potential, the pinning energy of such a crystalline density 
configuration depends on the choice of the computational lattice sites at
which the periodic local peaks of the density are located. 
We take, among all possible choices
equivalent by
symmetry operations in
the computational lattice, the one for which the pinning energy is lowest
for the specific pinning configuration under consideration.
These initial
conditions are used to obtain ordered states at lower temperatures. Finally,
states originally obtained  by either of these two procedures can
be slowly warmed up, or cooled down: in this case the initial
conditions are simply the final state obtained at the previous temperature.
As one warms up or cools down a state, its nature in general changes: liquid
configurations may become unstable upon cooling and ordered configurations
upon warming. 

Whichever initial conditions one uses, a local minimum of the free energy is
eventually found. At a given temperature, in general several local
minima with different density configurations, characterized by
the values of the set of $\{\rho_i\}$ variables, are obtained. 
The minimum with
the lowest free energy represents, at a given $T$, the true equilibrium state.
It is obviously necessary to develop systematic procedures to identify
the structure of each of
these diverse minima. Since we have access to the full set of density
variables at the computational lattice sites, we have at our disposal a variety
of methods to achieve that goal. The first is to calculate the density
correlations, e.g. the structure factor $S({\bf k})$:
\begin{equation}
S({\bf k})=|\rho({\bf k})|^2/N_v
\label{sofk}
\end{equation}
where $\rho({\bf k})$ is the discrete Fourier 
transform of the set $\{\rho_i\}$. Equivalently, one
can consider the Fourier transform of $S({\bf k})$,
which is the two-point
spatial correlation  function $g({\bf r})$ of the time-averaged local
density. We will present here results 
for $S({\bf k})$, considered as a function of the vector
${\bf k}$, and for the angularly averaged spatial density correlation
function, $g(r)$. 

It is also very useful to consider the spatial
structure formed
by the {\it vortices} at low temperatures.
For this purpose we have extracted from $\{\rho_i\}$ the local peak
densities. We say that 
the density locally peaks at site $i$ if the value of $\rho_j$ at $j=i$ 
is higher than that at any other computational lattice site within a distance
$a/2$ from $i$, where, as defined above, $a$ is the equilibrium spacing
of the unpinned vortex lattice. As expected, we find that at low-temperature
solid-like minima where the vortices are strongly localized, 
the number of local density
peaks matches the number of vortices $N_v$.
The positions of these peaks determine what we will call
the ``vortex lattice''.

Much useful information about the spatial structure of this
vortex lattice  may be obtained by plotting directly 
the local density values at the
vortex lattice points. An excellent, and complementary, alternative to 
help elucidate the
degree of vortex lattice order is to carry out a Voronoi construction:
we recall that in the Voronoi construction one determines cells around each
lattice point by a Wigner-Seitz procedure. 
In a perfect crystal, all the
resulting Wigner-Seitz cells
are identical, while in the general case the cells can have
different sizes and shapes.
The number of sides of the  Wigner-Seitz cell surrounding a lattice point
is identified with the number 
of ``nearest neighbors''
of that lattice point, and the difference between this number and the average
(six in our case) marks the position of defects (disclinations). We will
make use of such plots below.  

Since we will be concerned not only about translational,
but also with orientational order, we will also examine the bond-orientational
correlation function $g_6(r)$ in the vortex lattice defined as:
\begin{subequations}
\label{angular}
\begin{equation}
g_6(r)=\langle\psi({\bf r})\psi(0)\rangle
\label{ang1}
\end{equation}
where the angular brackets denote overall average over the vortex
lattice and the field $\psi({\bf r})$ is given by:
\begin{equation}
\psi({\bf r})=\frac{1}{n_n} \sum_{j=1}^{n_n} \exp[6i\theta_j({\bf r})]
\end{equation}
\end{subequations}
with $\theta_j({\bf r})$ being the angle that the bond connecting
a vortex at ${\bf r}$ to its $j$-th neighbor makes with a fixed
axis, and $n_n$ is the number of neighbors of the vortex at ${\bf r}$.

It is also possible, and as we shall see, useful, to define what we will
call a ``translational correlation function'' in the vortex lattice in a way
that is quite analogous to the definition of $g_6(r)$. This function,
which we denote as $g_G(r)$, is defined  by an equation identical
to the right-hand side of Eq.~(\ref{ang1}), but with the field $\psi$
being defined as
\begin{equation} 
\psi_G({\bf r})=\exp(i {\bf G} \cdot {\bf r})
\label{bo}
\end{equation}
where ${\bf G}$ is a reciprocal lattice vector of the triangular vortex lattice
in the absence of pinning. We will consider here
only the case where ${\bf G}$ is a shortest nonzero reciprocal lattice
vector and average over the results obtained for the six equivalent $\bf G$'s.

\begin{figure}
\includegraphics [scale=0.6] {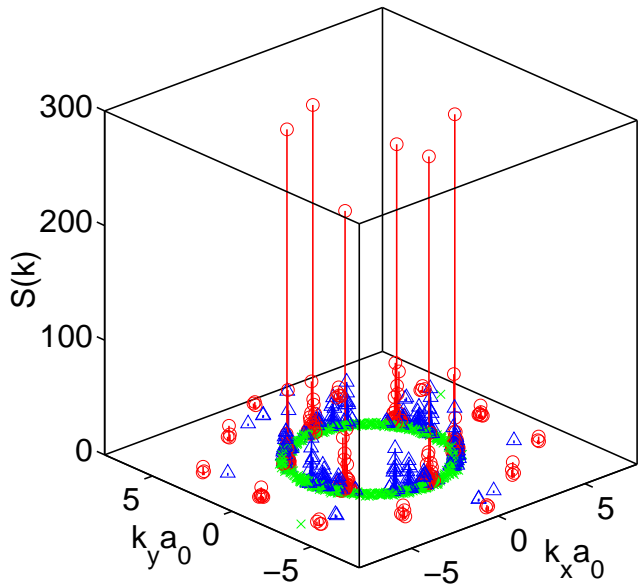}
\caption{\label{fig1}(Color online) The static structure factor  $S({\bf k})$
(see Eq.~(\ref{sofk})) for three local minima obtained at $T=18.4$K for
the same pin configuration. The (red) circles are for a state
identified, from the overall evidence (see text), as a Bragg glass (BrG), 
the (blue) 
triangles are for 
a Bose Glass (BoG) state, and the (green) crosses are for an insterstitial
liquid (IL). The vertical lines
are guides to the eye.}
\end{figure}

Our identification of the different phases is based on analysis of all 
this information. We show some
of the results in  the next few Figures. These all correspond
to  samples with 64 pins and 4096 vortices.
First, in Fig.~\ref{fig1}, we show
the structure factor, as defined in Eq.~(\ref{sofk}). The three sets of results
shown there are for local free energy minima at the same temperature, $T=18.4$K,
obtained with different initial conditions of the three kinds
described above, and the same pin configuration. Other pin 
configurations yield very similar 
results (a different example is shown in Fig.~1c of Ref.~\onlinecite{dv3}).
The green symbols correspond to the free energy minimum obtained by starting
with uniform initial conditions. Clearly, the structure factor is completely
featureless and liquid-like in this case -- the absolute value of $S$
never exceeds five. The red circles are for a local minimum
obtained  with initial conditions corresponding 
to the best crystalline state, 
as explained above, for this pin configuration.  The structure factor
now exhibits typical ordered behavior, highlighted by six sharp Bragg-like
peaks which are emphasized by the added vertical lines in the Figure. 
Finally, the blue triangles correspond to a minimum obtained by first
``quenching'' with uniform initial conditions to a relatively low
temperature (16.8K) where the liquid-like state is 
found not to be stable, and then
slowly warming, at 0.2K temperature intervals, back up to 18.4K. As one
can see, the $S({\bf k})$ for this minimum has an 
intermediate structure, with several
relatively well-defined peaks, more than six in number, whose heights are
considerably lower than those of the peaks found for the ordered minimum.

\begin{figure}
\includegraphics[scale=0.7] {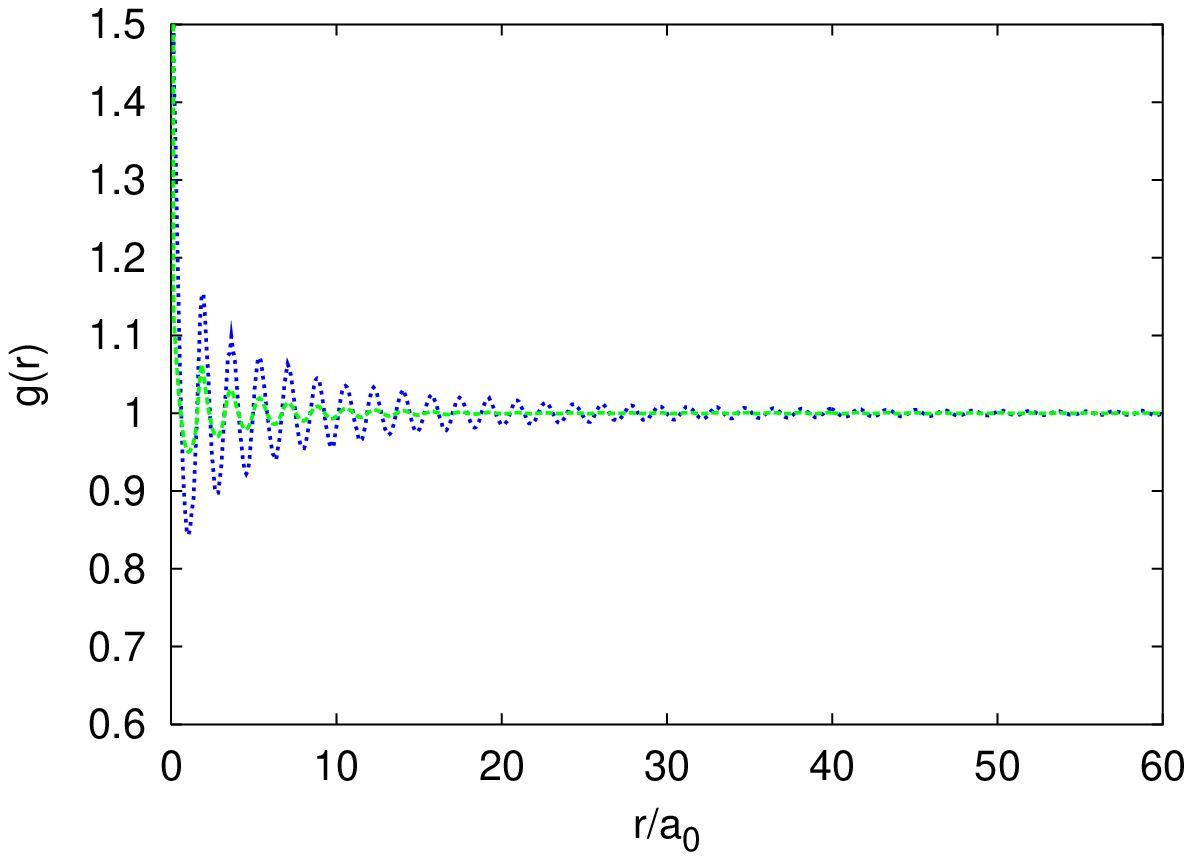}
\includegraphics[scale=0.7] {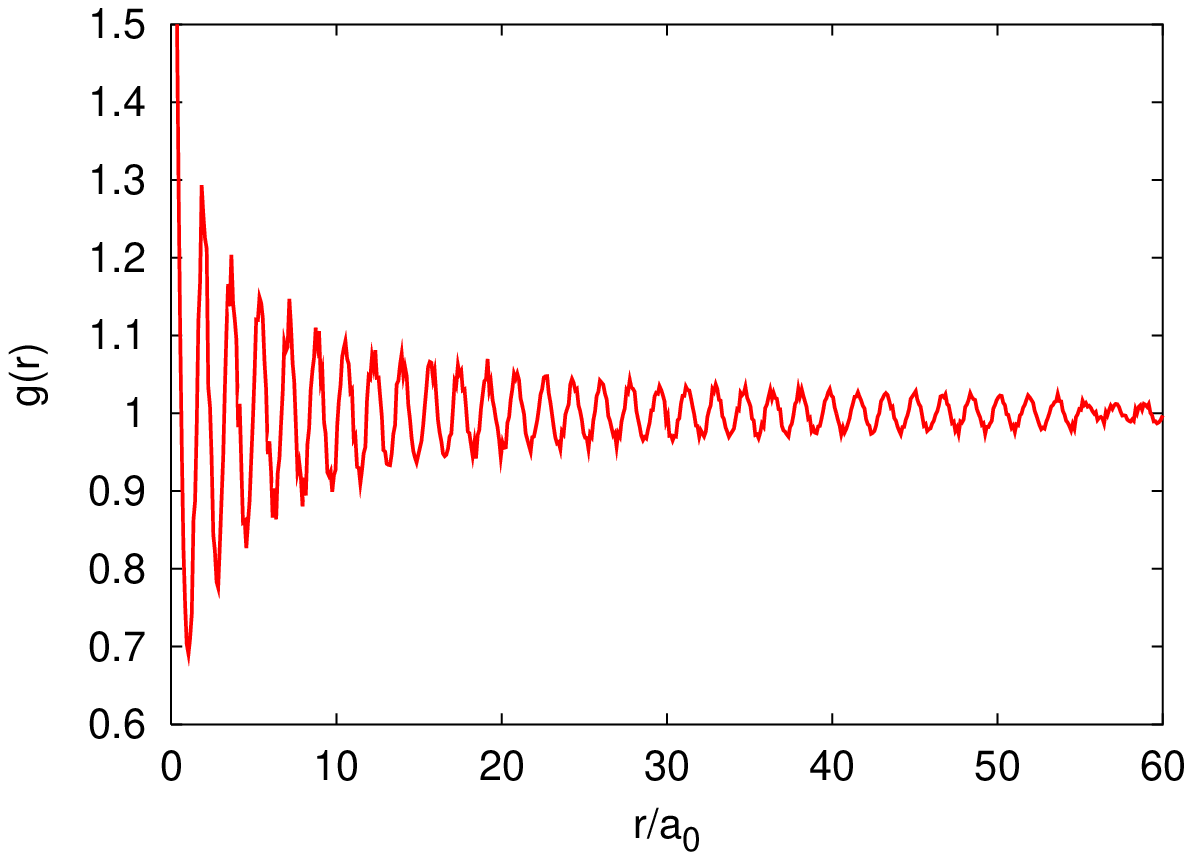}
\caption{\label{fig2} (Color online). The angularly averaged density
correlation function $g(r)$ plotted at the same temperature
as in Fig.~\ref{fig1}
as a function of dimensionless distance. The results are this time averaged
over five different pin configurations. The results for the BoG and IL
minima are shown in the top panel. The (blue) dotted
line plot with higher peaks corresponds
to BoG minima and the (green) light grey plot with fewer and lower 
peaks 
to IL minima. The bottom panel displays the results for the BrG
minima. The vertical and horizontal scales are the same in the two
panels and the color scheme is the same as that in Fig.~\ref{fig1}.} 
\end{figure}

In Fig.~\ref{fig2}, results for the angularly averaged density correlation
function $g(r)$ are displayed. These
results are for minima obtained at the same temperature, and 
using the same initial condition procedures, as those for
the data shown in Fig.~\ref{fig1}. However, the results shown here
are averages over five different pin configurations.
The nearly flat (green) curve in the top panel corresponds to minima obtained
from quenches with uniform initial conditions, the
(red) curve with the well-defined
peaks shown in the bottom panel 
corresponds to minima obtained with 
crystalline initial conditions, and the (blue)
curve with the intermediate peaks shown in the top panel is for
minima obtained by quenching with uniform initial conditions 
to a low temperature and 
subsequent slow warming (very similar results can 
alternatively be obtained by slowly cooling a high-temperature liquid-like
minimum to a temperature where it is unstable). 
One can see then that the results
for $g(r)$ are fully consistent with those found in Fig.~\ref{fig1}: the 
minimum obtained from uniform initial conditions is fully disordered, while
that obtained from crystalline initial conditions exhibits a large degree of
crystalline order.
For the the third kind of minima, we find some degree of intermediate
range order.

\begin{figure}
\includegraphics [scale=0.40]{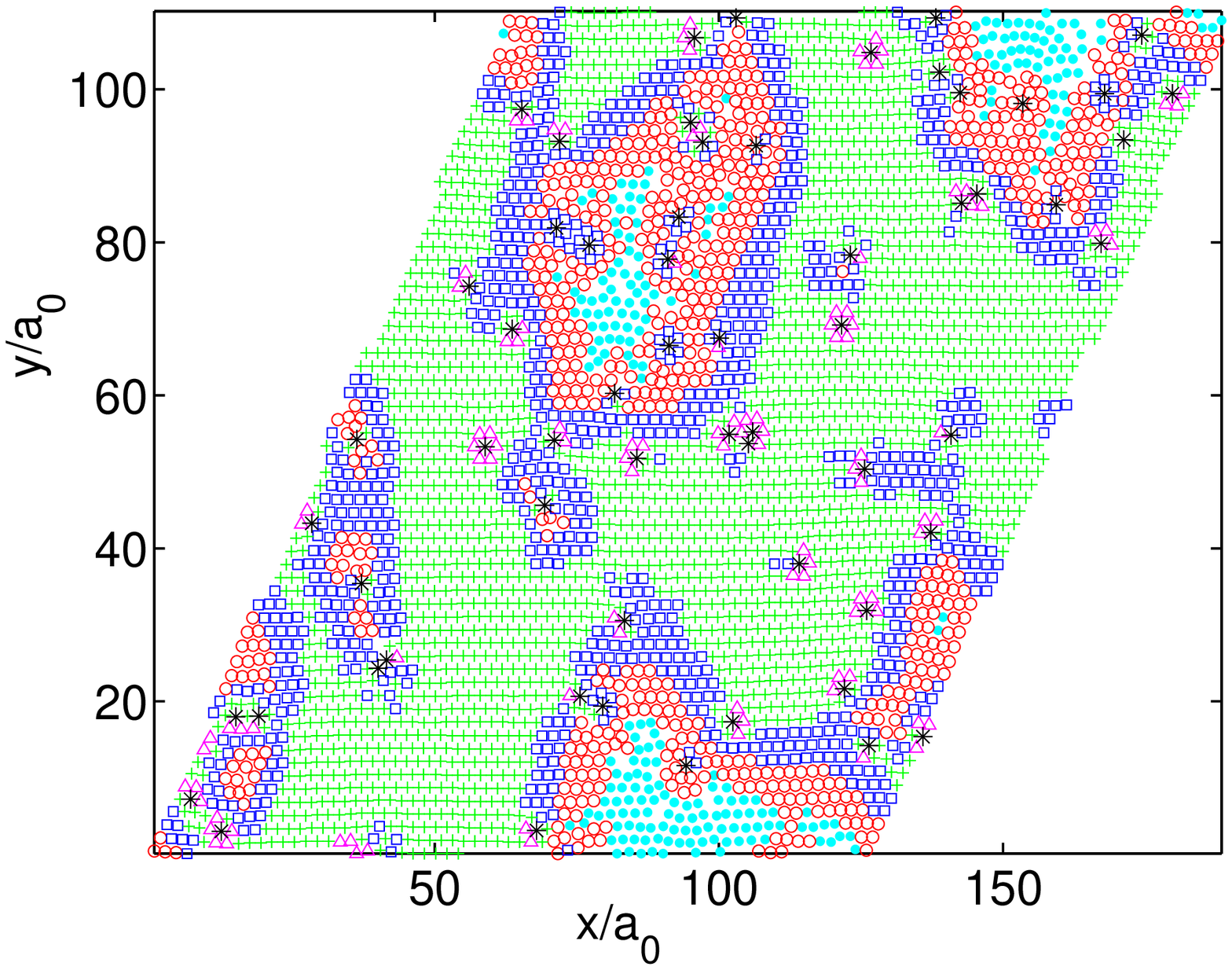}
\includegraphics [scale=0.40]{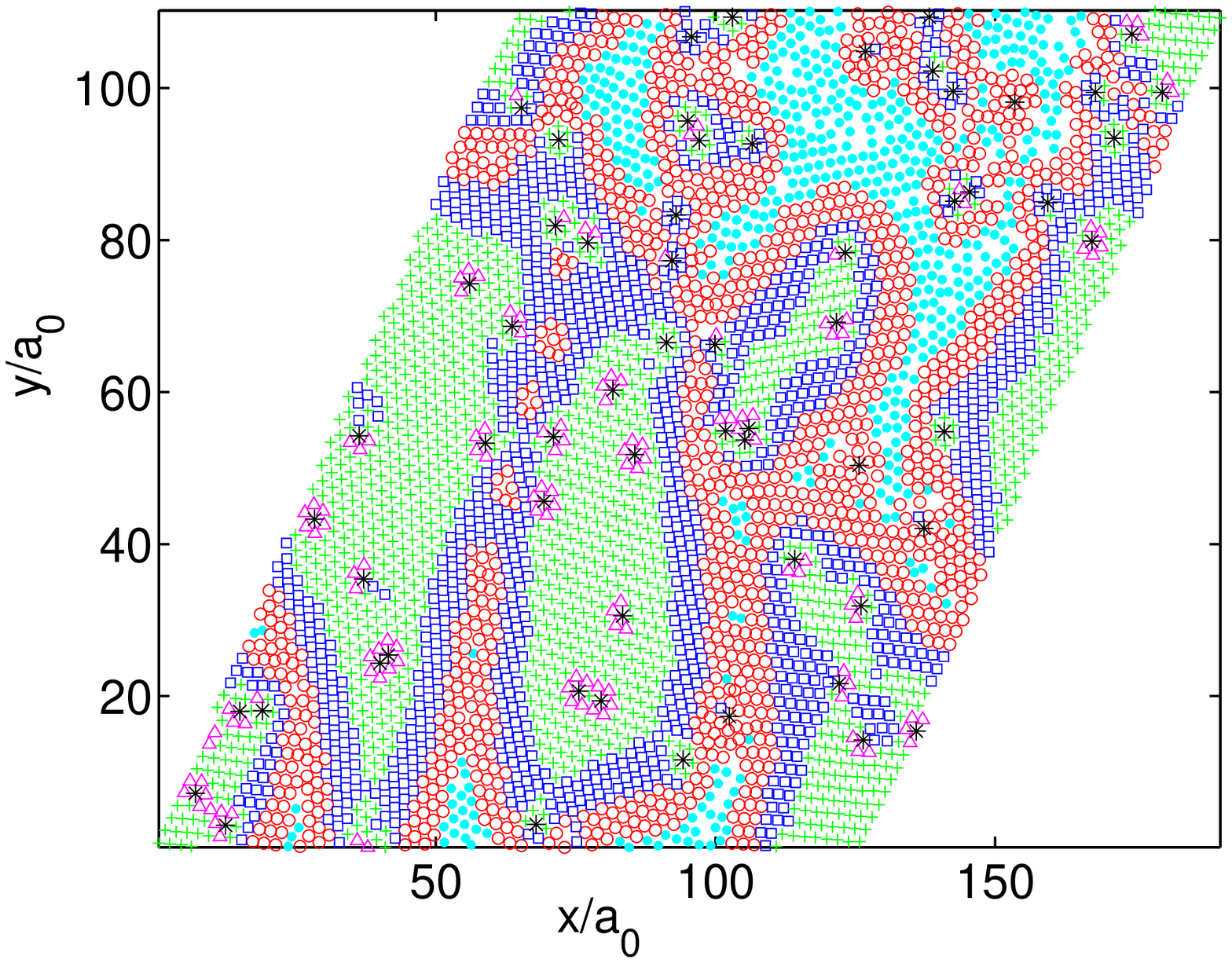}
\includegraphics [scale=0.40]{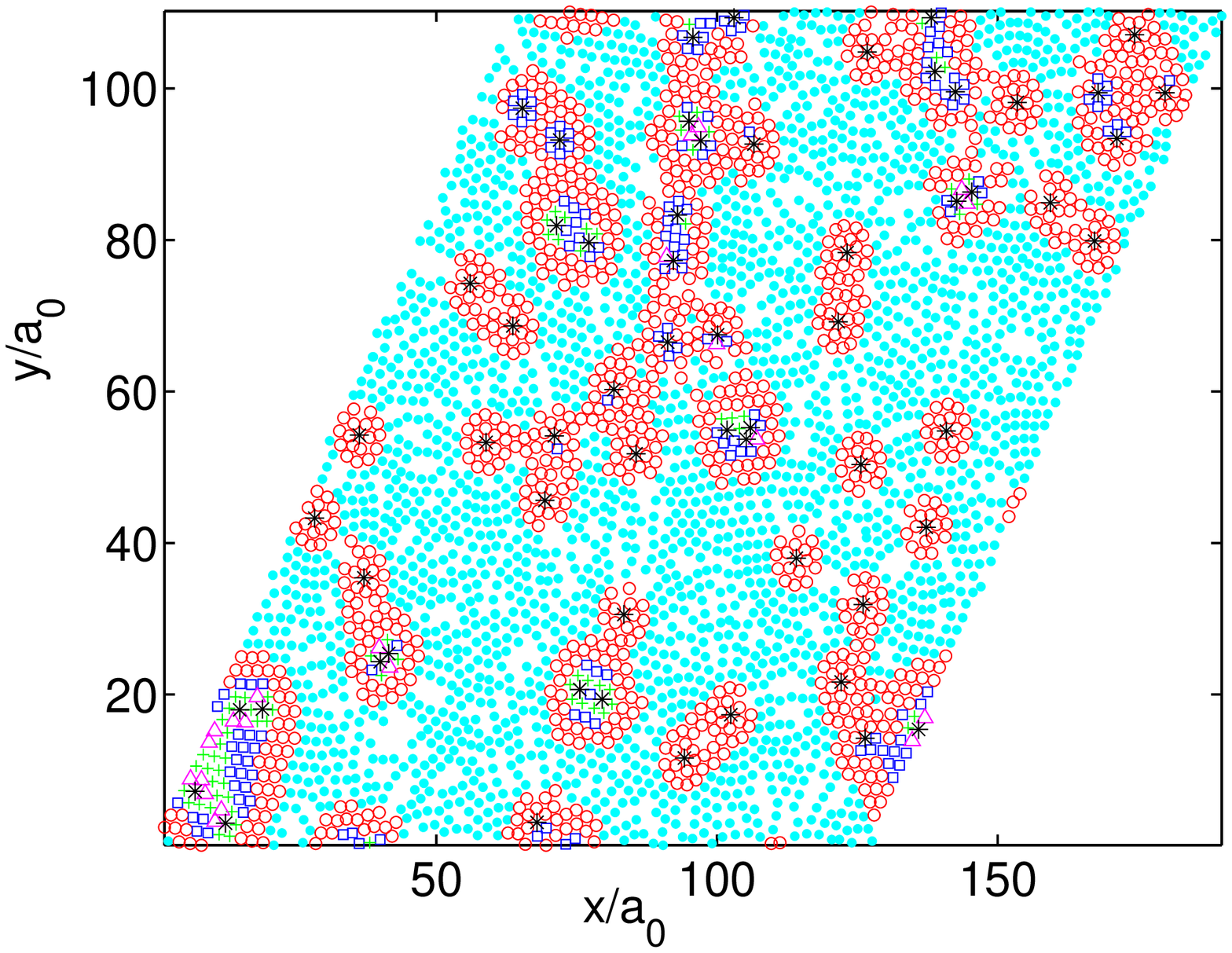}
\caption{\label{fig3} (Color online). Local peak densities at three
different local minima of the free energy. The positions of the
local density peaks and the values of the density at these peaks 
are displayed according
to the following scheme: (cyan) solid circles:
peaks with $\rho_{\hbox{peak}}/\rho_0
< 1.5$, (red) circles: $1.5 \le \rho_{\hbox{peak}}/\rho_0 < 3.5$, (blue)
squares: $3.5 \le \rho_{\hbox{peak}}/\rho_0 < 5.5$,
(green) plus signs: $5.5 \le \rho_{\hbox{peak}}/\rho_0 < 7.5$, and
(purple) triangles: $7.5 \le \rho_{\hbox{peak}}/\rho_0 < 9.5$. Blank
areas, found mainly in the bottom panel, correspond
to regions where no local peaks are found. The
temperature is 18.2K  in all panels, which all correspond
to the same pin configuration  as that for
the results in Fig.\ref{fig1}. Pin locations
are indicated by (black)
asterisks. From top to bottom, the results displayed
correspond
respectively to ordered (identified
as BrG), intermediate (BoG) and disordered (IL)
minima. }
\end{figure} 

We next examine the structure of the minima in terms of the ``vortex lattice''
found by the procedure explained above. Sample results are shown in 
Fig.~\ref{fig3}, all for the same pin configuration, at the temperature
$T=18.2$K. Results for three minima, obtained from the same procedures
and initial conditions as the three minima in the previous two figures,
are shown. The positions of the local  density peaks and the 
values of the density at these peaks are displayed through a symbol
and color coding scheme described in the caption of the Figure. 
Blank regions denote areas where no local peak was found using the algorithm
described above: such
areas are mostly 
found only in the most disordered case (uniform initial conditions).
We see that the results are quite consistent with those obtained from
the previous correlation functions: quenching with the appropriate
crystalline initial conditions leads to a fairly well-ordered (but not
perfectly ordered) lattice, while from uniform initial conditions, one
obtains a rather disordered, liquid-like structure where the density
is nearly uniform and close to $\rho_0$, except near the pinning centers
at each of which one vortex is trapped.  An intermediate result, on
the whole more solid than liquid-like, but definitely disordered, is
obtained through the slow warming or cooling scheme.

\begin{figure}
\includegraphics [scale=0.45]{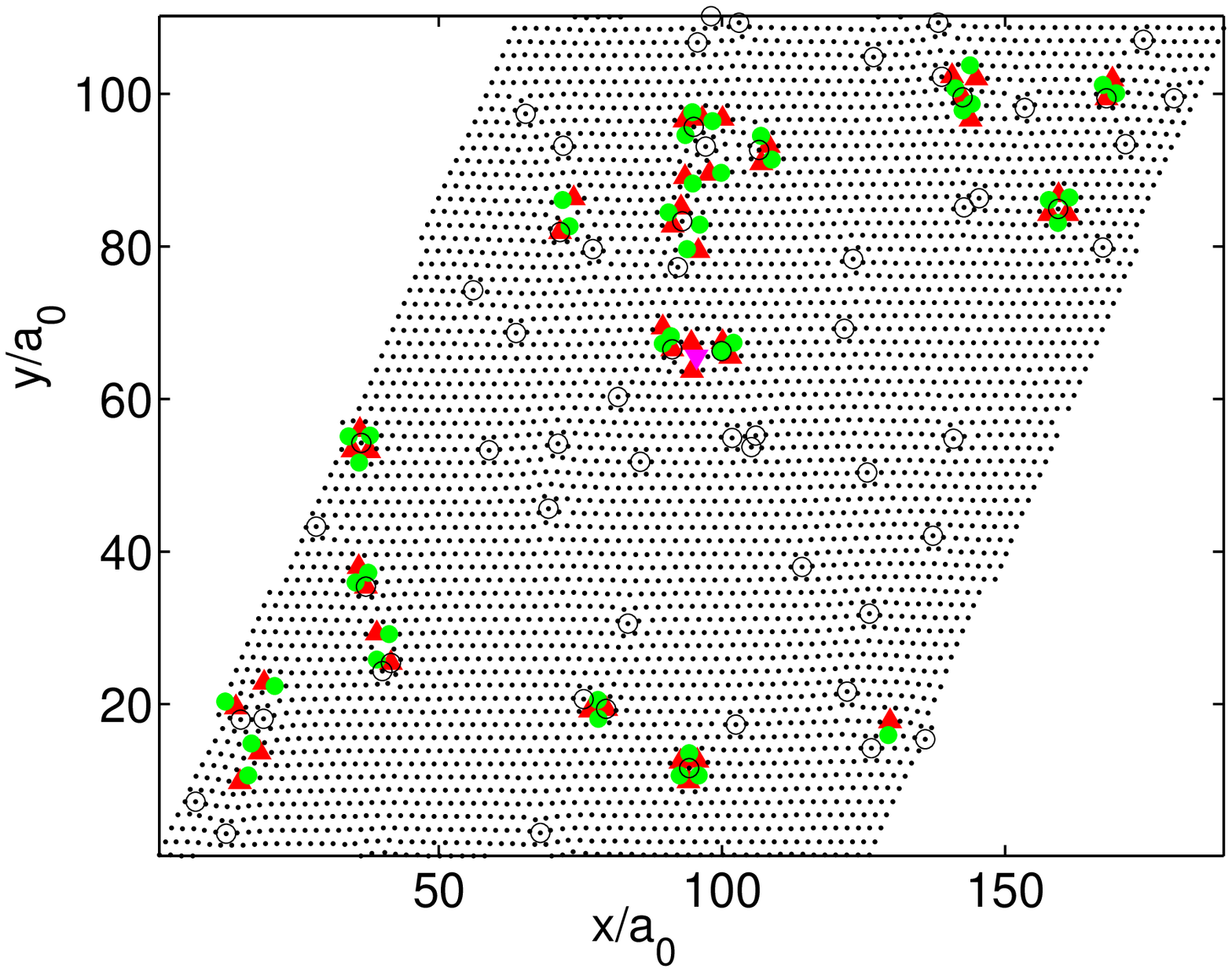}
\includegraphics [scale=0.45]{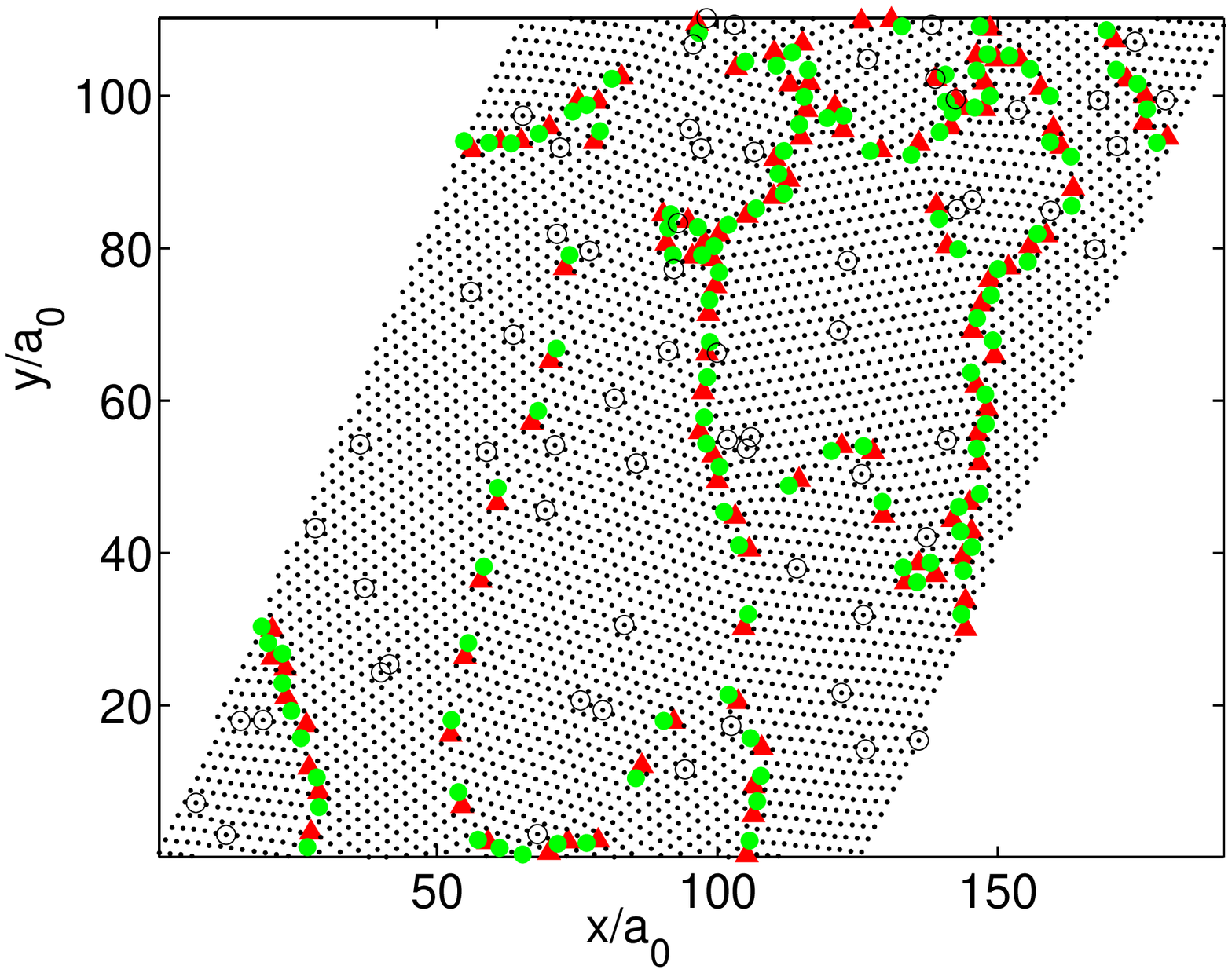}
\includegraphics [scale=0.45]{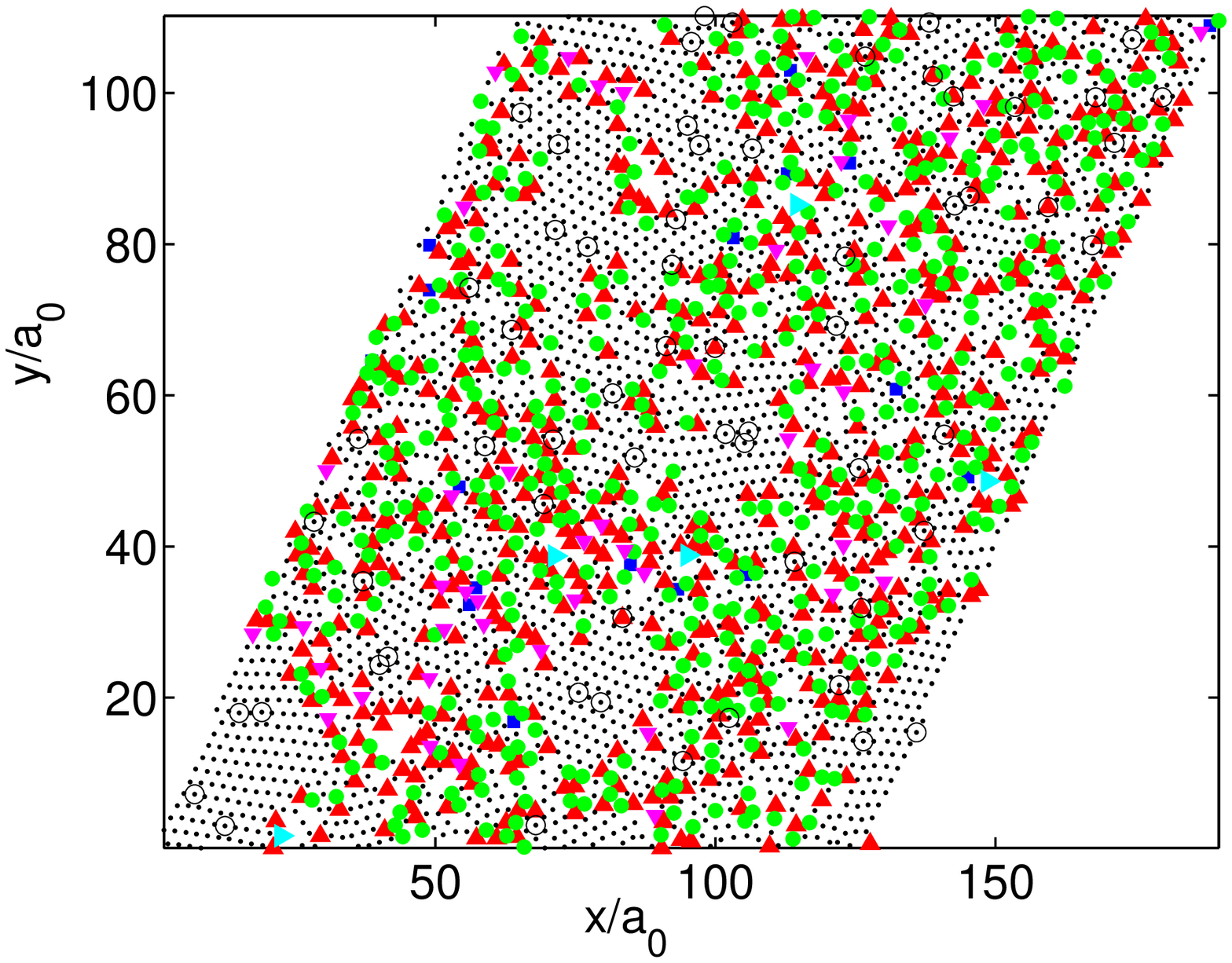}
\caption{\label{fig4} (Color online). Voronoi plots for the three cases
for which density plots are shown in Fig.~\ref{fig3}, except that
the temperature is now $T=18.0$K. The (black) dots denote
lattice sites with six neighbors, the (red) triangles denote
five-fold coordinated sites, and the (green) solid circles,
seven-fold coordinated sites. Rarely occurring four-fold and 
eight-fold coordinated
sites are indicated by (blue) squares and (purple) inverted triangles,
respectively. Sites
surrounded by black circles denote locations of pinning centers. See text
for explanations.}
\end{figure}

To better gauge the degree of disorder present in each case, we construct,
in Fig.~\ref{fig4}, the corresponding Voronoi plots. To display the behavior
in a more obvious way, the temperature in these plots is lower by $0.2$K than
that of the plots in Fig.~\ref{fig3}. As mentioned above, sites with number of
nearest neighbors ($n_n$) different from six represent the locations of
disclinations. In the solid-like minima (top two panels of Fig.~\ref{fig4}),
nearly all the sites with $n_n \ne 6$ have $n_n=5$ or $n_n=7$ (the 
disclinations have unit ``charge''). Also, six- and seven-coordinates sites,
indicated respectively by (red) triangles and (green) solid circles in the 
plots, always
occur in nearest-neighbor pairs. Such pairs correspond to dislocations.
In the first panel, which displays the
results for the most ordered state obtained from crystalline initial
configurations, 
we see that the number of dislocations, while nonzero, is quite 
small and that they form fairly isolated  small
clusters, each of which has zero net Burgers vector. These defect
clusters occur near the pinning centers and they represent the local
disruption of crystalline order due to the pinning. 
In the second panel, which shows the results for the minimum with
intermediate order, the dislocations 
are organized to form well-defined 
grain boundaries that separate crystallites with different orientations. 
As a result of this polycrystalline
structure, both translational and bond-orientational correlations are 
short-range. Performing the same construction
for other pin configurations (results for a different pin configuration 
are shown in Fig.1b of Ref.~\onlinecite{dv3}), 
one finds that the crystallite arrangement
depends on the pin configuration, as one
would expect. The grain boundaries tend to
lie away from pinning centers, which makes sense physically.
Finally, in the last panel, which shows the results for the minimum 
obtained from uniform initial conditions, the high degree
of disorder is evident. The Voronoi construction
shown in this panel is not very meaningful
because our method of obtaining the vortex positions
is not reliable for liquid-like minima.
The identification of the position
of a vortex with that of a local peak of the density field is justified
only when the peak is sharp. This is the case in the solid-like minima, but
not so in the liquid-like minima with low peak densities. We have shown here
the results for the liquid-like minimum only for the purpose of 
illustrating the difference between the structure of this minimum and those
of the more ordered ones. For this
minimum, we find a very large
number of defects, 
and it is difficult to determine whether
disclinations of opposite ``charge'' always pair up to form dislocations. 
Inspection
of the defect distribution suggests the presence of free disclinations,
but not conclusively. The total number of
local peaks of the density field in liquid-like minima turns out to be
substantially smaller than the expected number of vortices.
This problem is not present at the solid-like minima 
where nearly all the vortices are
strongly localized. 

\begin{figure}
\includegraphics [scale=0.45] {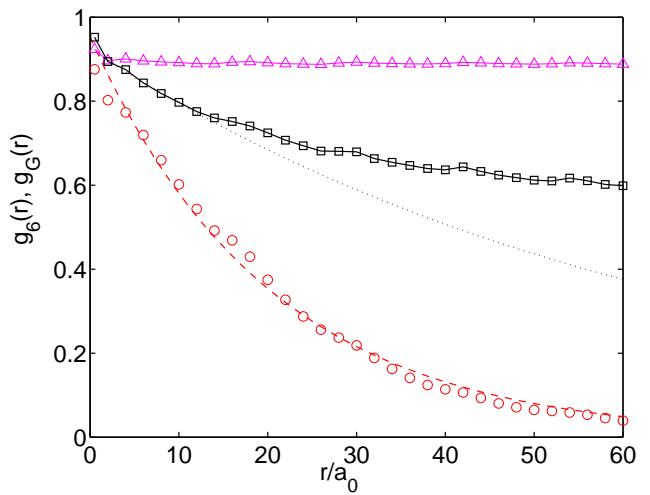}
\caption{\label{fig5} (Color online). Bond-orientational  correlation function
$g_6(r)$ as defined in Eq.~(\ref{angular}), and ``translational'' correlation
function, $g_G(r)$, see Eq.~(\ref{bo}). These are results averaged over
five pin configurations, at $T=17.6$K. The (purple) triangles and (red)
circles denote $g_6(r)$ for ordered (BrG) and polycrystalline (BoG)
minima (liquid-like
minima are not stable at this lower $T$) respectively.  The black squares
are $g_G(r)$ for the BrG minima. The dashed line is an exponential
fit to $g_6(r)$ for the BoG minima. 
The dotted line is a similar exponential fit to
$g_G(r)$ for the BrG minima 
at small distances, showing that the decay is slower than 
exponential for large values of $r$. 
The solid lines connecting symbols are guides to the eye. }
\end{figure}

Next, we show in Fig.~\ref{fig5} examples of the
bond-orientational and ``translational'' correlation functions, $g_6(r)$
and $g_G(r)$ as
defined in Eqs.~(\ref{angular}) and (\ref{bo})
respectively. All results shown in this
figure are averages over five pin configurations 
at  temperature $T=17.6$K. At this
temperature the liquid is unstable, and in any case all liquid correlation
functions are featureless, so that case is not shown.
The (purple) triangles correspond to ordered minima obtained from quenching
with crystalline
initial conditions to $T=16.8$K and subsequent slow warming, and represent
the bond-orientational correlation, $g_6(r)$. This function appears to 
saturate at a fairly large value as $r$ increases, indicating the presence
of long-range bond-orientational order. The 
(red) open circles display $g_6(r)$ for the polycrystalline 
state with intermediate order,
obtained through the procedures described above. This function decays 
exponentially with distance, as shown by the exponential fit
(dashed line), over a distance of few tens in units of $a_0$. Finally, the 
(black) open squares show the results for
the ``translational'' correlation 
function $g_G(r)$ for the ordered minima. 
The dotted line is an exponential fit to this translational 
correlation function for $r/a_0 < 10$, showing that the decay at
longer distances is clearly slower 
than exponential (possibly a power-law).

>From the overall examination of data such as those shown in these  
figures, we can reach the following conclusions: quenching with uniform
initial conditions to temperatures above or slightly below $T_m^0$ leads
to very disordered states, with very short range correlations and no
structure, except for vortices at and near the pinning centers. These minima
become unstable upon slow cooling: one then obtains minima of the third kind,
discussed below. They clearly must be identified with the liquid phase,
specifically the interstitial liquid (IL) phase discussed in the Introduction.
The second kind of minima are obtained by quenching with crystalline initial
conditions to any temperature below, or slightly above, $T_m^0$. These minima
can then be cooled, without losing their character, but they melt
into the IL upon sufficient warming. The states thus obtained
are obviously nearly crystalline, with the reservation that the order is
not truly long range, but has a slow decay. This is most evident in the
results for $g_G(r)$ seen in Fig.~\ref{fig5}. The bond-orientational order
is nearly perfect. Defects are limited to isolated clusters. This state
must therefore be identified as a Bragg glass (BrG). Finally, we have the states
obtained as described above, by slow cooling of the IL state,
which can also be obtained by quenching
with uniform initial conditions to a temperature below the IL state
stability limit. Either procedure yields, for any pin configuration,
states that at a given $T$ differ only slightly in free energy
or density configuration. The states do differ somewhat more, however, for
different pin configurations. The Voronoi construction conclusively
shows (see Fig.~\ref{fig4}) that these states are polycrystalline. The
grain boundaries are formed by dislocation chains. For any pin configuration,
these grain boundaries survive thermal cycling across $T_m^0$, 
but if sufficiently
warmed up, these minima melt, the melting beginning locally at the 
grain boundaries (see Sec~\ref{props} below).
The bond-orientational correlation function for these minima 
has an intermediate range. This,
and also the observation 
that these states are not unique (as mentioned above, slightly
different states are reached depending on the cooling or quenching protocol),
clearly indicate a glassy state. We therefore identify it with the
Bose glass (BoG) which has been experimentally\cite{ban,meng} and
numerically\cite{sen} shown to
be polycrystalline.
 
\subsection{Properties of the condensed phases}
\label{props}

In this section, we describe in detail how the properties of the low-temperature
``solid-like'' condensed phases represented by the BrG and BoG types of minima
defined in the preceding section vary as the temperature is changed. The 
high-temperature liquid phase represented by the IL-type minima is not very
interesting from the point of view of its temperature dependence: in the
temperature range we have considered, this phase exhibits a
liquid-like (nearly
uniform) density distribution except in the immediate vicinity of the columnar
pins, at each of which a vortex is trapped. 
These trapped vortices would eventually get delocalized
at higher temperatures. This  happens beyond the upper
limit of the  temperature
range (16.8K $\le T \le$ 18.8K) we have considered here.

To study how the properties of a local minimum of the free energy change as the
temperature is varied, we have ``followed'' minima as the temperature
changes: for example, starting with a minimum obtained by
quenching to the
lowest temperature  of 16.8K, we ``follow'' that minimum
to higher temperatures by increasing the temperature
in small steps (usually taken to be 0.2K) and finding a new minimum at the
higher temperature by running the minimization routine with the configuration
at the minimum obtained at the previous temperature
as the initial state. This procedure leads to a new
minimum of the same type as long as the minimum remains locally stable --  if
the temperature is increased to values 
substantially higher than $T_m^0$, where the BrG and BoG minima
become unstable, this procedure leads then to the IL minima. We have also
carried out  ``cooling'' runs where the temperature was decreased in
small steps, starting from a minimum obtained at a relatively high temperature.
When the IL minima  become unstable, BoG states are obtained.
As long as the minimum under consideration 
did not become unstable in the range of temperatures considered in the runs,
we did not find any substantial difference between the results obtained in the
``heating'' and ''cooling'' runs

\begin{figure}
\includegraphics [scale=0.7] {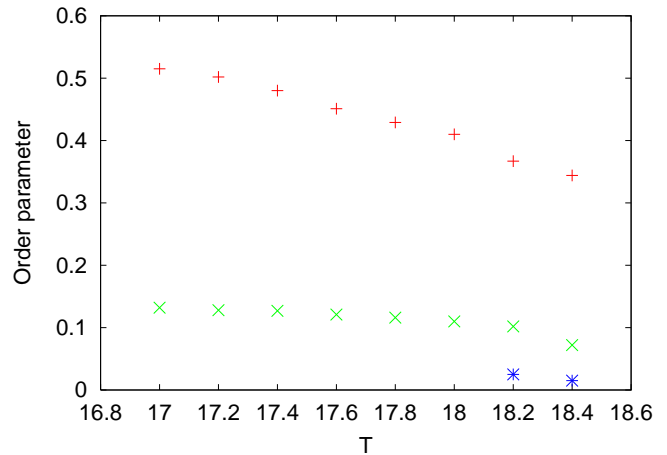}
\caption{\label{fig6} (Color online). Temperature dependence of the order
parameter $m$  as defined
in Eq.~(\ref{op}) for three kinds of minima, averaged over five samples, with 
$N_v=4096$ and $N_p=64$. Results for the BrG, BoG and IL minima are shown
by (red) plus signs, (green) crosses and (blue) asterisks, respectively.}  
\end{figure}

To characterize the density distribution at a minimum, we looked at ``global''
quantities such as the structure factor $S({\bf k})$, an example
of which is shown above in Fig.~\ref{fig1}.  The $T$-dependent
information 
contained in 
$S({\bf k})$ is, however, difficult to display in plots.
 We
therefore introduce  a closely related
``order parameter'' $m$ defined as
\begin{equation}
m \equiv \sqrt{(S_{max}/N_v)}
\label{op}
\end{equation}
where $S_{max}$ is the largest value of the structure factor $S({\bf k})$ 
averaged over the six $\bf k$-vectors related by 
lattice symmetry to each $\bf k$. 
By definition, the order parameter $m$ is equal to unity in a state with
perfect crystalline order (triangular lattice with $\delta$-function peaks). 
Since no long-range
crystalline order is expected or found in either one of the BrG and BoG phases, the
value of $m$ should go to zero at all temperatures in the thermodynamic
limit. However, the  infinite size limit is reached
very slowly in the glassy phases, and
for the finite-size systems considered here, this quantity
provides a convenient measure of the degree of local order present at the
minima.

We show in Fig.~\ref{fig6} the results for the order parameter $m$, averaged
over five different pin configurations,  for samples with 4096
vortices and 64 pins. The results for the BrG, BoG and IL minima are shown
by (red) plus signs, (green) crosses and (blue) asterisks, respectively.
Only two data points are shown for the IL minimum because it becomes
unstable at lower temperatures. For all three kinds of minima, the value of
$m$ decreases with increasing $T$, as expected. 
The rate of change  is largest for the BrG
minima, and smallest for the IL minima. However, the differences
among the three kinds of minima in the degree of local order remain quite clear
at all the temperatures considered. We have also obtained BrG-type minima
for several ($\sim 10$) samples with 1024 vortices and 16 pins. The values
of $m$ obtained from an average over these smaller samples are found to be
only about 6\% larger than those obtained for the 4096-vortex samples
at the corresponding temperature. This result implies that the translational
correlation function at the BrG minima falls off very slowly with distance.
This is consistent with our interpretation of these minima as
representing the BrG phase.

\begin{figure}
\includegraphics [scale=0.45]{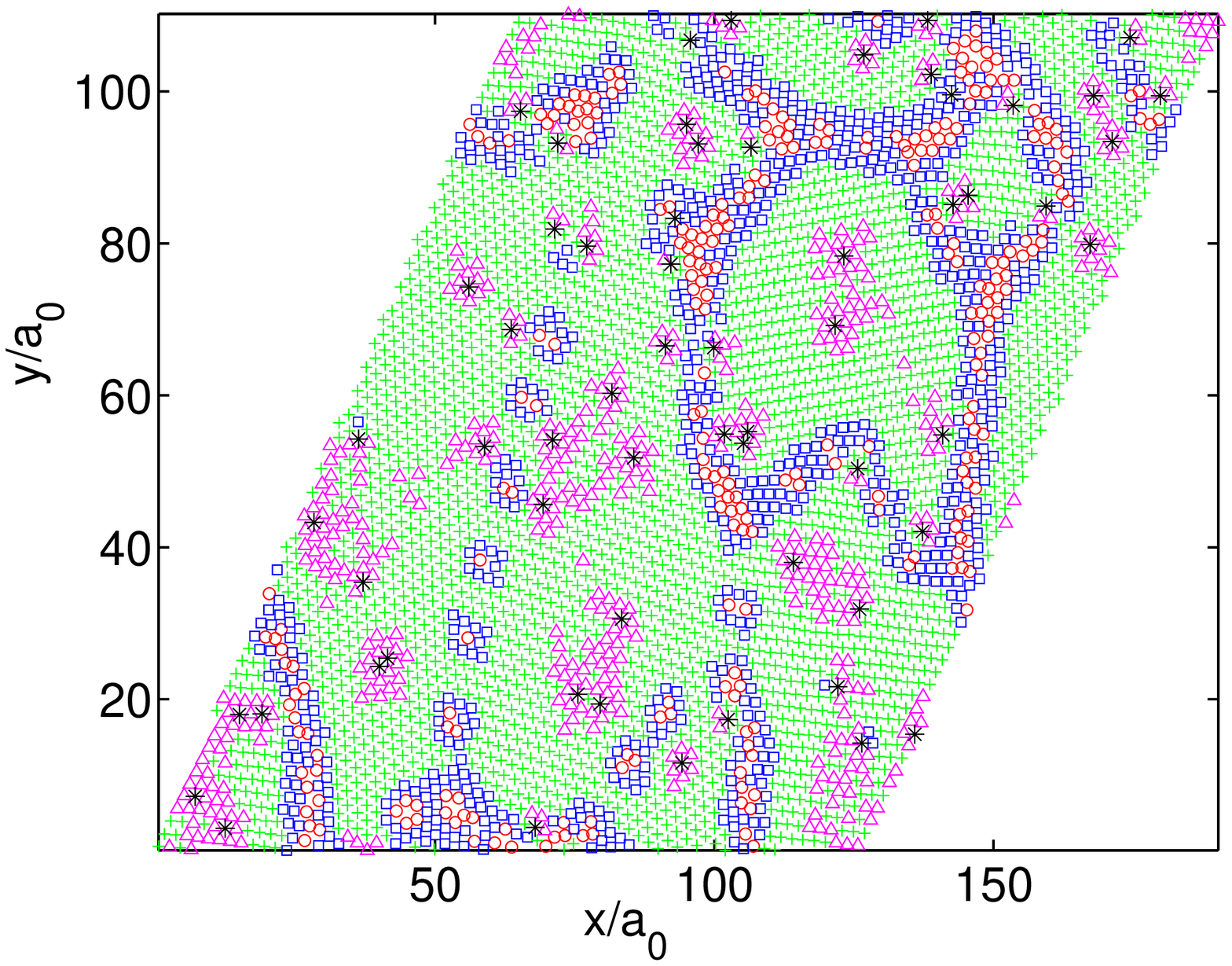}
\includegraphics [scale=0.45]{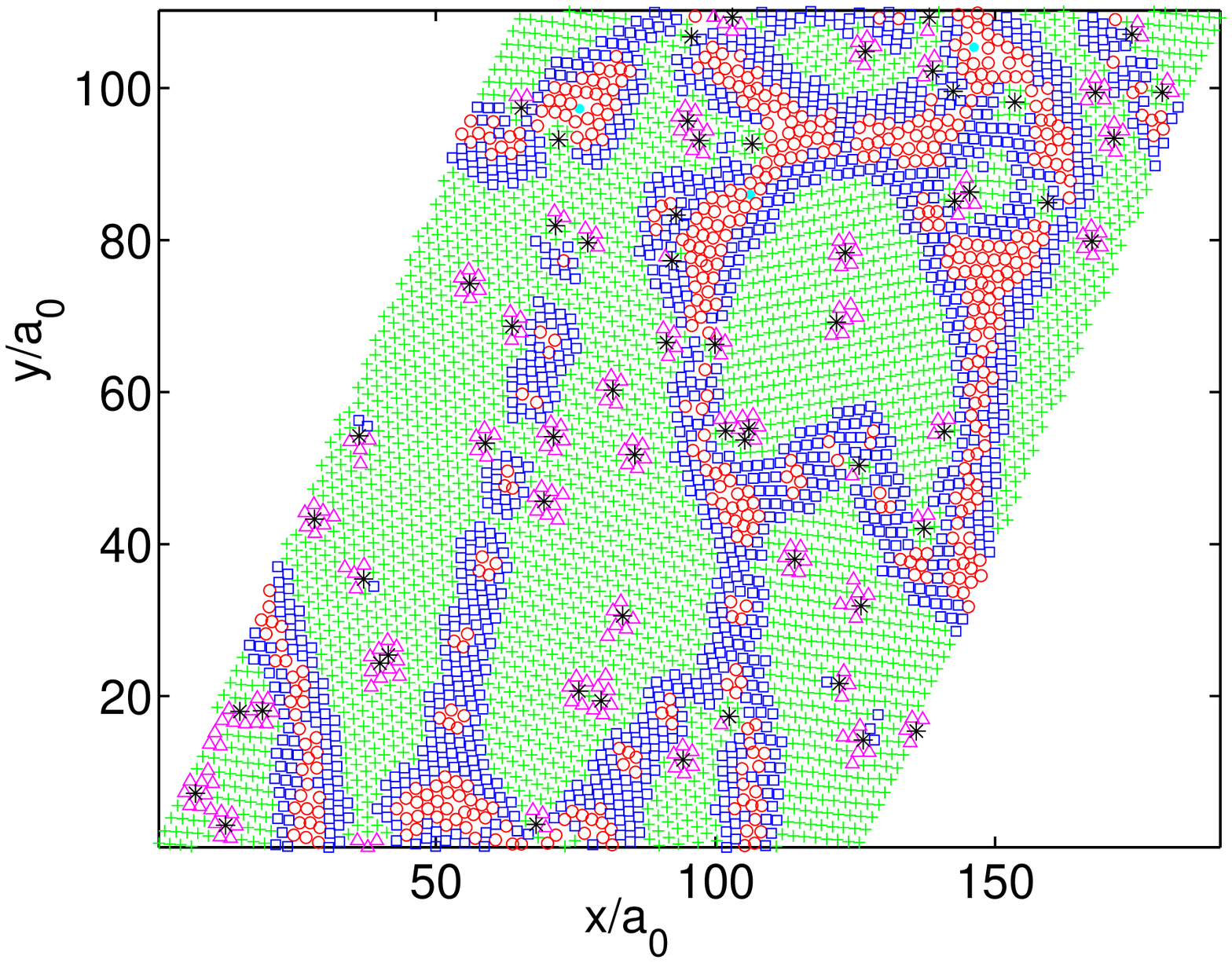}
\caption{\label{fig7} (Color online). Peak density plots 
for the BoG minimum of a sample with $N_v=4096$ and $N_p=64$ (the sample
is the same as the one for which results are shown in Figs.~\ref{fig3}
and \ref{fig4}) at two temperatures: $T$ = 17.4K (top panel) and 
$T$ = 17.8K (bottom panel). The symbols and color scheme used here are the
same as those in Fig.~\ref{fig3}.}
\end{figure}

We also examined the temperature dependence of the detailed density
distribution at the minima (obtained from ``peak density''
plots similar to those
shown in Fig.~\ref{fig3}) and their defect structure obtained from Voronoi
plots such as those in Fig.~\ref{fig4}. 
Examination of the peak-density plots reveals more information about how
the density distribution at the minima changes with temperature. In 
Fig.~\ref{fig7}, we have shown peak-density plots at two temperatures (17.4K
in the top panel and 17.8K in the bottom panel) for the BoG minimum of the
same sample for which a similar plot at 18.2K is shown in the
middle panel of Fig.~\ref{fig3}. The symbols and color scheme used in these
plots are the same as those in Fig.~\ref{fig3}. It is clear from these plots
that the typical values of the local peak densities decrease (the vortices
become less localized) as the temperature  increases. The largest values of
the local peak density occur near the pinning centers and the smallest
values appear near the grain boundaries that separate different crystalline
regions. This is physically reasonable -- the disruption of local crystalline
order near the grain boundaries should make the vortices in such regions more
delocalized. At the relatively low temperature of 17.4K, there are very few
local density peaks where the peak density is lower than $3.5 \rho_0$ (such
peaks are indicated by 
(red) circles in the plot). As the temperature is
increased to 17.8K, the number of such peaks increases, and this trend
continues as $T$ is increased further, as can be seen in the
middle panel of Fig.~\ref{fig3} where the results for $T=18.2K$ are shown.
It is also clear from these plots that the spatial regions where the low
peak densities occur remain roughly unchanged as the temperature 
increases -- as noted above, these regions are strongly correlated with the 
locations of the grain boundaries. 
These observations indicate that the vortices in the interior of the
crystalline grains remain in a ``solid'' state as the temperature is 
increased to a value close to $T_m^0$, while those in the neighborhood of
the grain boundaries  begin to ``melt'' (get
delocalized) at a lower temperature. This ``inhomogeneity'' of the melting
process will be discussed in more detain in section~\ref{inhomog}.

We also studied the temperature dependence of the number of topological
defects present at the minima. This number increases slowly with increasing
$T$, and then exhibits a sudden jump as liquid-like regions (characterized
by low values of the local peak densities) appear near the grain boundaries
at temperatures close to $T_m^0$. As noted in Sec.~\ref{minima} above,
our method of obtaining the defect structure
using the Voronoi construction becomes somewhat
less reliable when liquid-like
regions begin to appear. 
Due to this difficulty in obtaining reliable results
for the defect structure and statistics at relatively high temperatures, we
have not carried out a quantitative analysis of the dependence of these
quantities on the temperature.

The gradual decrease in the values of the local peak densities with increasing
temperature is also found in the BrG minima, but on the average
the vortices remain more
strongly localized at the BrG minima than at the corresponding
BoG ones. The strong correlation between 
low values of the local peak density and the location of topological defects
is found at the BrG minima too. The main difference between BrG
and BoG minima is that the defects are more randomly distributed at
the BrG minima: they do not line up along grain boundaries as they do at
the BoG minima. For this reason, the regions of low peak density appear 
at fairly random locations at the BrG minima. The total number of topological
defects at the BrG minima remains smaller than that at corresponding BoG
minima at all temperatures. As the temperature approaches $T_m^0$, liquid-like
regions appear in parts of the sample where the defect density is large.
This is illustrated in the plots in the top panels of Figs.~\ref{fig3} and
\ref{fig4}.

Thus, our investigation of the temperature
dependence of the structure of the BrG and BoG minima
confirms that these minima retain their topological structure as 
$T$ is increased
toward $T_m^0$ and even beyond (provided they remain
stable). The degree of localization of the vortices decreases with
increasing $T$ and liquid-like regions with nearly delocalized vortices
begin to appear at temperatures close to $T_m^0$. However, even at such
temperatures, the BrG and BoG type minima are clearly distinguishable from
one another and from the IL-type minima, as illustrated in Figs~\ref{fig3},
\ref{fig4} and \ref{fig6}.

\subsection{Two-step melting transition}
\label{phasediag}

As discussed in the preceding subsections,
we find  three kinds of 
coexisting, locally stable minima of the free energy 
at temperatures close to $T_m^0$.
Two kinds (BrG and BoG) of local minima are found at temperatures 
substantially lower than $T_m^0$, and only the IL minimum is stable if $T$
is much higher than $T_m^0$.
In our mean-field description,
the thermodynamically stable phase at a particular temperature 
corresponds to the minimum with the lowest free energy. 
Therefore, crossings
of the free energies of different kinds of 
minima correspond to first-order phase
transitions in our description. The IL minima are expected to have the lowest
free energy at high temperatures and the solid-like BrG or BoG minima should 
represent the  globally stable phase at low temperatures. To determine how
this ``freezing'' transition takes place, it is necessary to examine the
dependence of the free energies of these different kinds of minima on the
temperature $T$.

\begin{figure}
\includegraphics [scale=0.75]{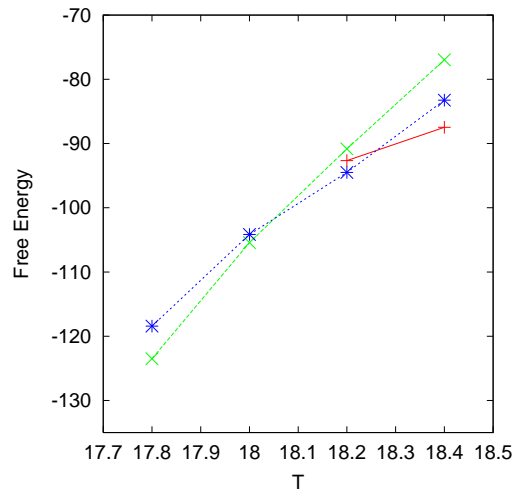}
\caption{\label{fig8} (Color online). Temperature dependence of the 
dimensionless free energies of different local minima of 
a sample with $N_v=4096$ and $N_p=64$ (the sample
is the same as the one for which results are shown in 
Figs.~\ref{fig1}, \ref{fig3},
\ref{fig4} and \ref{fig7}). The data for the BrG, BoG and IL minima are 
shown by the
(green) crosses and dashed line, (blue) asterisks and dotted line and the 
(red) plus signs and solid line,
respectively. The lines are guides to the eye.}
\end{figure}

In Fig.~\ref{fig8}, we have shown the results for the free energies of different
minima of a sample with $N_v=4096$ and $N_p=64$. This sample is the same as the
one for which results are shown in Figs.~\ref{fig1}, \ref{fig3}, \ref{fig4},
and \ref{fig7}. A similar plot, showing the same
general behavior for a different sample, may be found in 
Ref.~\onlinecite{dv3}.
 The BrG minimum clearly has the lowest free energy at low
temperatures (while we have shown data for $T \ge 17.8K$, we have checked
that this remains true at lower temperatures), while the IL minimum is the
one with the lowest free energy at high temperatures. The interesting feature
that emerges from data as that
shown in this Figure is that the BoG minimum has
the lowest free energy in an intermediate temperature range of small width
(between 18.0K and 18.3K for this sample), indicating that the melting of the
low-temperature solid phase to the high-temperature IL phase occurs in two
distinct steps: the BrG phase that is the thermodynamically stable one at low
temperatures undergoes a first-order transition into a BoG phase as the 
temperature is increased, and then
this BoG phase melts into the IL phase via a second first-order 
transition at a slightly higher temperature. This two-step melting behavior is
one of our main results. 

\begin{figure}
\includegraphics [scale=0.40]{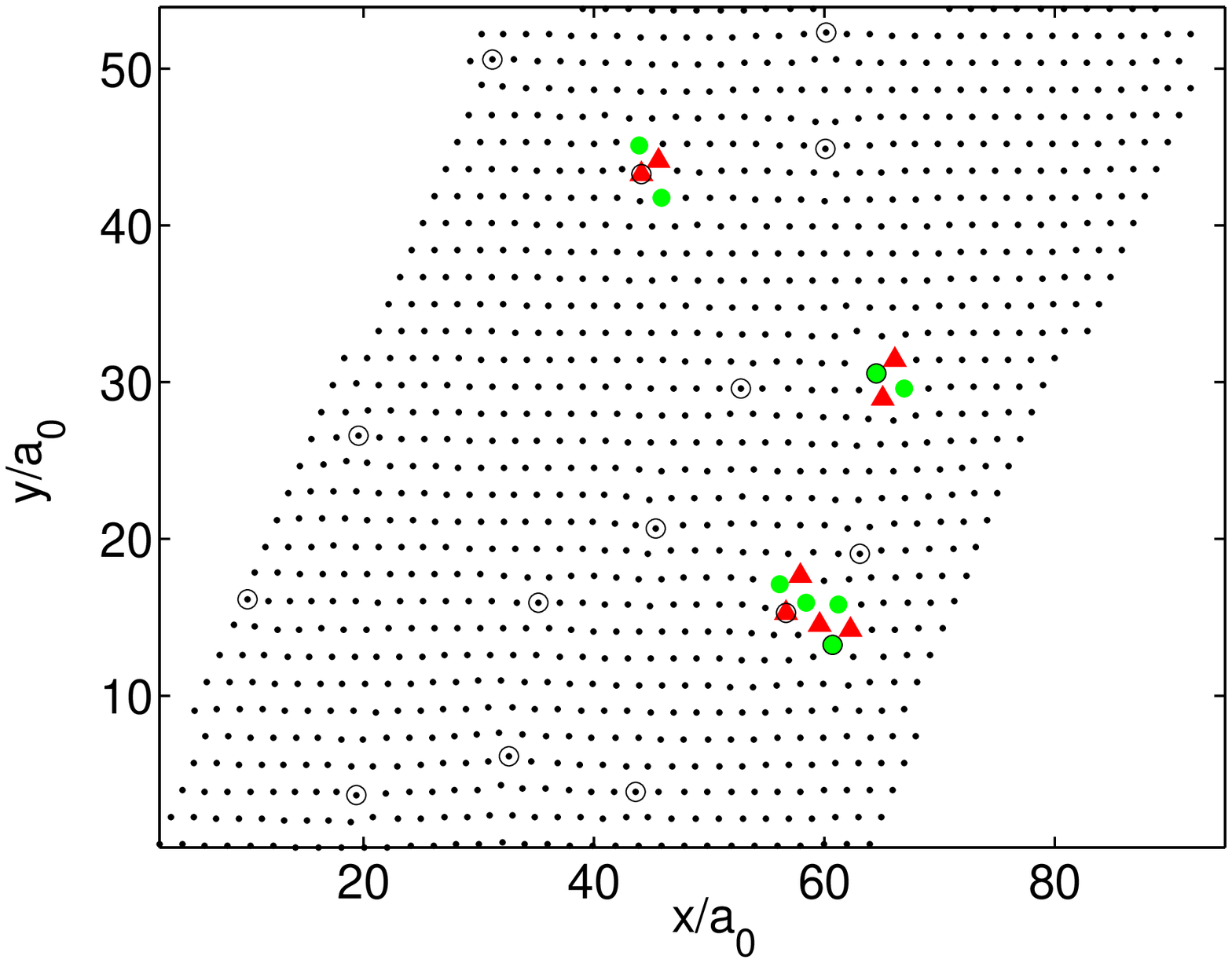}
\includegraphics [scale=0.40]{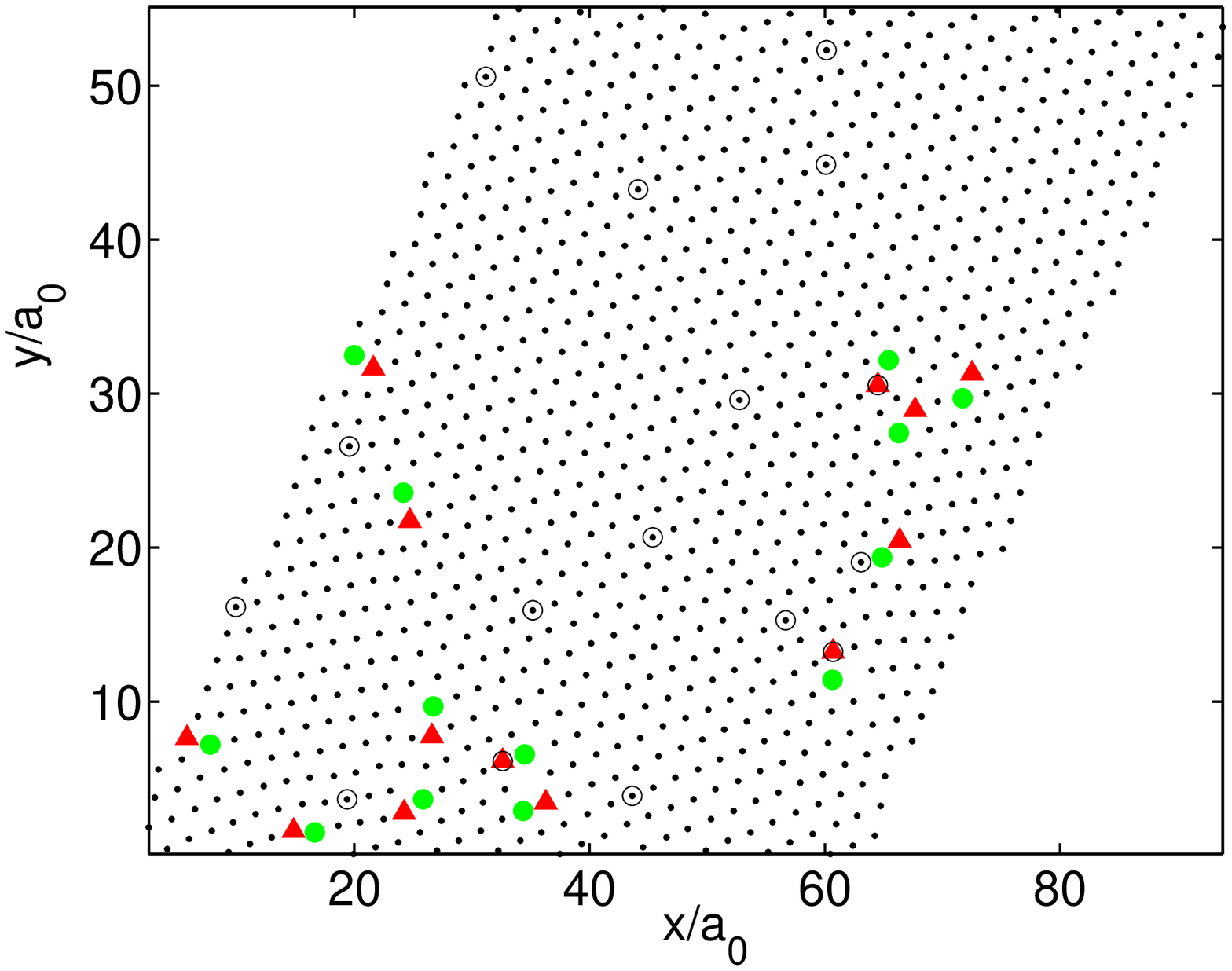}
\includegraphics [scale=0.40]{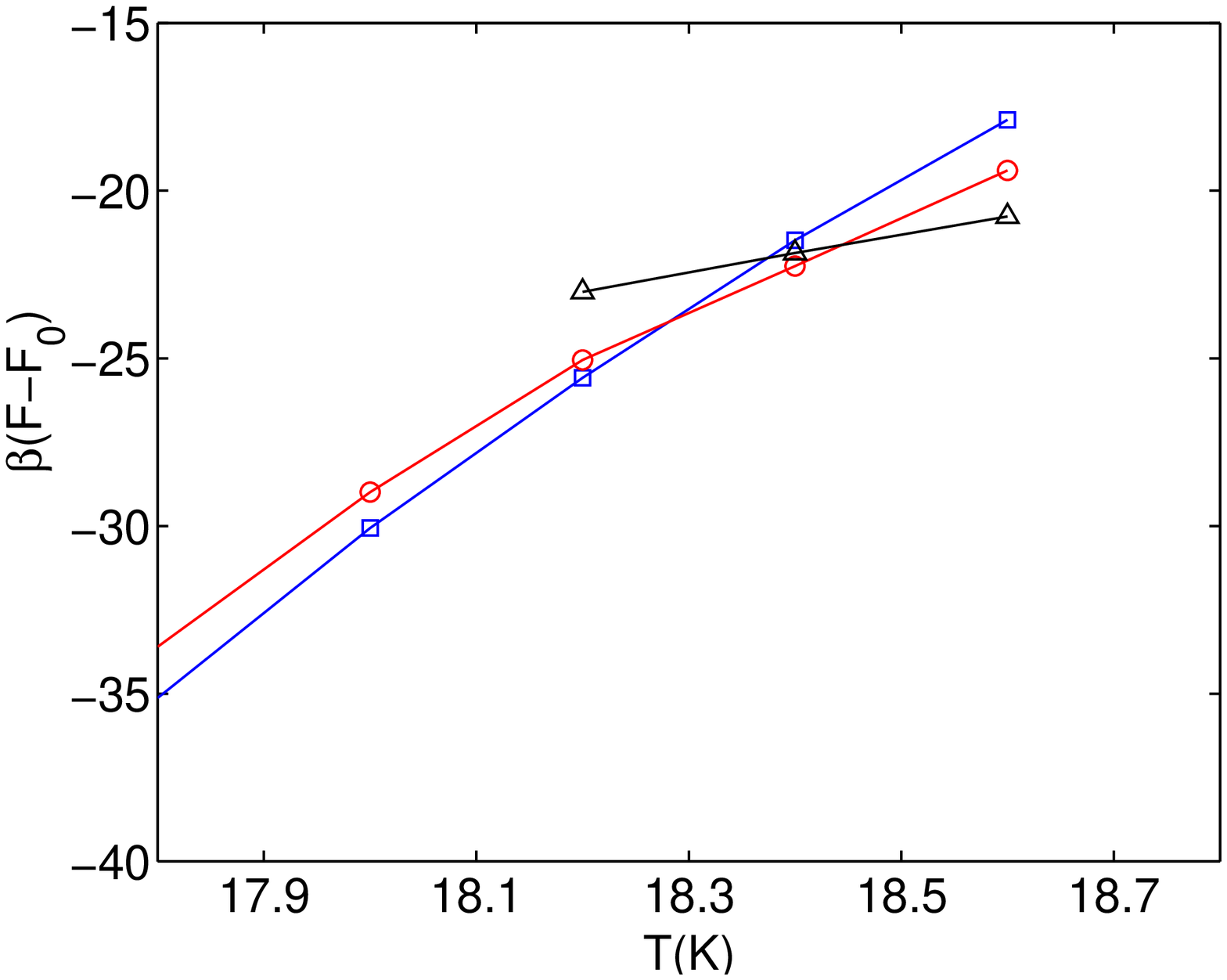}
\caption{\label{fig9} (Color online). Results for a sample with $N_v = 1024$
and $N_p=16$. Voronoi plots for the BrG and BoG minima at 18.0K are shown in
the top and middle panels, respectively. The symbols and color scheme used in
these plots are the same as those in Fig.~\ref{fig4}. The bottom panel shows
the temperature dependence of the free energies of the BrG (blue line and
squares), BoG (red line and circles) and IL (black line and triangles) minima,
respectively. The lines are guides to the eye.
}
\end{figure}

The same qualitative behavior, indicating the occurrence of two separate
first-order transitions, is found in all the 4096-vortex samples we have 
studied at pin concentration $c=1/64$ ($N_p=64$). The average
value of the temperature interval in which the BoG phase has the lowest free
energy is about 0.42K. This width exhibits fairly large sample-to-sample
variations, ranging between about 0.1K in one sample to a maximum of 
1.2K. As the system is cooled from the 
high-temperature liquid phase, it undergoes a first-order transition into
a polycrystalline BoG phase. Such a transition has been observed~\cite{meng} 
in experiments on BSCCO samples with a small concentration of columnar pins.
The value of the upper (BoG to IL) transition lies between 18.2K and
18.3K in all the 4096-pin samples we have studied. 
These values are quite close to the
first-order melting temperature $T_m^0$ of the same system in the absence
of pins~\cite{dv2}. Our results, thus, are consistent with the experimental
observation~\cite{kha,ban} of a weak dependence of the freezing temperature
of the vortex liquid on the pin concentration $c$ for small values of $c$.
In addition, our work predicts a second first-order transition to a more ordered
BrG phase at a slightly lower temperature. 

For smaller samples at the same $c=1/64$ concentration ($N_v=1024$ with
$N_p=16$) we did not
always find the BoG minima: the minimization procedure
that led to the BoG minima in the samples with $N_v=4096$ often converged to
minima of the BrG type in the smaller samples. This is because
the typical size of the crystalline grains at the BoG minima at this pin
concentration is of the order of the 
sample size for 1024-vortex samples. 
This can be seen by comparing  the middle panel
of Fig.~\ref{fig9}, where we have shown the Voronoi plot for
the BoG minimum obtained for a $N_v=1024$ sample,
with the Voronoi plot in the middle
panel of Fig.~\ref{fig4}, which is for the
same $T$ and $c$, but a larger sample with 4096 vortices. One can plainly see
that the domain size is comparable to the system size of the smaller
samples. 
However, for the
1024-vortex samples where we found BoG-type minima, the behavior of the free
energies was found to be very similar to
that shown in Fig.~\ref{fig8}. An example of such behavior is shown in 
the bottom panel of Fig,~\ref{fig9}. 
The two-step melting transition found in
the larger systems is found here also, indicating that this is the generic
behavior. We have also carried out preliminary studies of the phase diagram
for a slightly higher value of the pin concentration, $c=1/32$. We find the
same qualitative behavior, namely the occurrence of two-step melting, with a
larger difference between the two transition temperatures. We therefore
conclude that the two-step melting transition we have found is a 
characteristic feature of our model of 
layered superconductors with a small concentration
of columnar pins.

\begin{figure}
\includegraphics [scale=0.45]{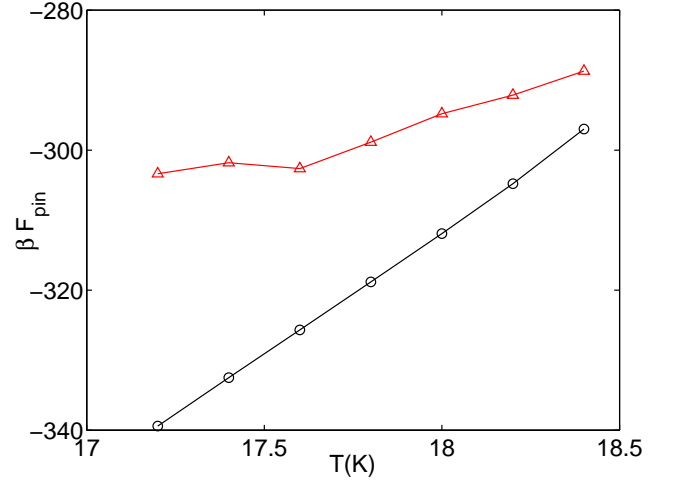}
\caption{\label{fig10} (Color online). The pinning  free
energy (see text) averaged over five samples with $N_v=4096$ and $N_p=64$.
Results for the BrG and BoG minima are shown by (red) triangles and (black)
circles, respectively. The solid lines are straight lines joining
the points.
}
\end{figure}

Since a first-order transition is not characterized by power-law behaviors
of thermodynamic quantities with universal exponents, the usual finite-size
scaling analysis of numerical data for continuous transitions does not apply
to our work. However, we have examined the dependence of the magnitudes of
various discontinuous changes at the transitions on sample size. From 
the results for the free energy as a function of
the temperature, we have calculated the entropy jump per vortex at the
first-order transitions. The entropy change at the BrG-BoG (lower
temperature) transition is found to be $\simeq 0.1 k_B$ per vortex ($k_B$
is the Boltzmann constant), and the corresponding change at the BoG-IL 
(higher temperature) transition has a slightly higher value, 
$\simeq 0.15 k_B$ per vortex. The sum of these two entropy jumps is slightly
smaller than the entropy change ($\simeq 0.29 k_B$ per vortex) 
at the single first-order melting transition found~\cite{dv2} in the same
system in the absence of pinning. All these results are physically
reasonable. The difference between the degrees of order in the BrG and
BoG minima is smaller than that between the BoG and IL minima, suggesting
that the entropy jump at the lower transition should be smaller than that
at the upper one. Also, the low-temperature BoG phase of the system in the
presence of pinning is less ordered than the crystalline phase of the pure
system, and the IL phase is slightly more ordered than the vortex liquid
in the system without pinning (this is due to the local order of the vortices
near the pinning centers which persists in the IL phase). 
Therefore, the net change of entropy in 
going from the BoG phase to the IL phase should be smaller than the
entropy jump at the melting transition in the pure system. A comparison of
our results for  two sample sizes ($N_v = 4096$ and 1024) 
does not show any appreciable dependence of the entropy changes on the
sample size. We, therefore, conclude that there is no indication of any
significant change in our phase diagram as the sample size is increased.

Since our mean-field treatment ignores
the effects of fluctuations, it is important to address the question of
whether one or both of the transitions found would become continuous
if fluctuations were included.  However, there are very few examples of
such fluctuation-driven continuous transitions in three dimensions. 
In our calculations, the
effects of the electromagnetic 
interaction among vortices on different layers are included.
So, we expect our results regarding the nature of the transition to remain
valid if fluctuations were included. This conclusion is supported by the
result that our density functional 
theory provides a {\it quantitatively} a correct
account~\cite{dv2,seng,men1} 
of the first-order melting of the vortex lattice in the pure system.
Also, the transition in the presence of a small concentration of columnar
defects is known to be first-order both experimentally~\cite{kha,ban,meng} 
and from simulations~\cite{non}.

The appearance of a sliver of the BoG phase in our phase diagram may be
qualitatively understood as arising from a competition between the elastic
and pinning parts of the free energy. The elastic (free) energy plays a 
dominant role in the BrG minima: the  effect of the randomly located pinning
centers is accommodated in this structure by small displacements of the 
vortices from their ideal lattice positions towards the nearest pinning center.
Where the occupation of a pinning center by a vortex
would require a large
displacement of the vortex, the pinning center
is {\it not} fully occupied. For this reason, the number of pinning centers
occupied by vortices at a BrG minimum 
is always slightly lower than the total number of
pinning centers. At the corresponding BoG minima, on the other hand, the
pinning centers are always occupied (except in very rare cases where two 
pinning centers are located very close to each other). Since the BoG minima
are obtained from liquid-like initial conditions, the starting point has
fully occupied pinning centers, and vortices localized in 
small crystalline patches in their immediate vicinity 
(see, for example, the bottom panel of Fig.~\ref{fig3} which shows such
crystalline patches surrounding the pinning centers). The
crystalline orientation of vortices 
around a vortex-binding pin site depends on the 
local arrangement of the pins.\cite{daf} This orientation
is, in general, different in different regions of the sample. These
crystalline patches grow  as the temperature is reduced, and 
eventually meet one another at grain boundaries to form a 
polycrystalline BoG minimum. It is clear 
that the pinning centers are
better accommodated in this structure than in the corresponding BrG structure.
On the other hand, the creation of grain boundaries costs elastic energy.
Thus, the BoG minima are expected to have lower pinning energy but higher
elastic energy than the BrG ones for the same pin configuration. 
The elastic energy dominates over the pinning energy at low temperatures
in our low pin-concentration samples. This is why the BrG phase is
globally stable at low $T$. Both these components of the free
energy decrease in magnitude as the temperature is increased. 
The softening of the lattice near the melting transition causes
the elastic energy to decrease faster than the pinning energy. This makes
the pinning component of the free energy
more important than the elastic part 
near $T_m^0$, thereby making the total free energy of the
BoG minima (which, as discussed above, have lower pinning energy) lower
than that of the BoG minima. The BoG to IL transition at a slightly higher
temperature is driven by the usual entropic mechanism.

We can substantiate this qualitative explanation 
by evaluating the pinning
component of the free energy, as defined in Eq.~(\ref{pin}),
for the BrG and BoG minima. The results of our calculation, averaged over
five samples (each with 4096 vortices and 64 pins), are shown in Fig.~\ref{fig10}.
The pinning energy is negative for both kinds of minima, as expected.
The absolute value
of the pinning energy of the BoG minima shows a smooth 
(nearly linear) decrease as the temperature is increased. This is because both
the value of the parameter $\Gamma$ (see Eq.~(\ref{single})) that
determines the depth of the pining potential, and the height of the local
density peak at a pinning center decrease with increasing $T$. The plot for the
BrG minima shows  a similar but much slower behavior,
still monotonic or nearly so: it is not clear
whether the very shallow minimum near $T$ = 17.6K 
is significant. In some of the 
samples, the total number of occupied pining centers 
increases by a small amount near
this temperature, thereby reducing the value of the total
pinning energy. This may reflect a better accommodation of the pinning centers
by the lattice, which softens as the temperature is increased. 
This plot clearly
demonstrates that the pinning centers are better accommodated at the
BoG minima. The difference in the pinning energies of the two kinds of
minima decreases with increasing $T$, but the difference in the 
elastic component of the free energy decreases faster (this is clear from our
results for the total free energy) due to a softening of the elastic
constants with increasing temperature. The overall effect  is a 
crossing of the free energies of the two kinds of minima near the 
melting transition. 

Quantitatively, however, the values of the two parts of
the free energy and their dependence on $T$ are determined by
the material parameters of the superconductor and the properties of the 
pinning centers.
Since several
experimental studies\cite{kha,ban,meng} 
of the effects of irradiation-induced columnar pinning on the mixed state
of BSCCO exist in the literature,  
we have used parameter values appropriate for this system.
For other layered superconductors, the BoG phase might occur
over a wider or narrower (even vanishing) range.

\subsection{Inhomogeneous melting of the vortex solid}
\label{inhomog}

As mentioned in section~\ref{props}, the density
distribution in both BrG and BoG minima is very
inhomogeneous: at temperatures close to $T_m^0$
there are liquid-like regions,
characterized by low values of the local peak densities in parts of
the sample. In this subsection, we show that
this inhomogeneity leads to a spatial variation of a ``local melting 
temperature'',
defined below.
As mentioned in the Introduction, such spatial variations of a local
melting temperature 
have been deduced from measurements of the local magnetization
 in  
BSCCO samples with random point\cite{soi} and columnar
pinning\cite{ban}.

Since we are dealing with time averaged densities, it is clear that the value
of $\rho_i$ at a local peak of the density field provides a measure of
the degree of localization of the vortex whose average position corresponds
to the location of the density peak. A high (low) value of 
the local peak density implies strong (weak) localization. 
Smaller values of the local peak density imply mobile, liquid-like
behavior. 
The value of the local density is, of course,
equal to $\rho_0$ everywhere in the liquid state in the absence of pinning.
The presence of pinning centers causes the local density to vary in space, but
in the liquid state this variation does not lead to local peaks higher 
than about $3\rho_0$ 
(excluding
the vortices localized at the pinning centers).
A similar result was also obtained in previous
studies~\cite{dv2} of vortices in the
presence of pinning, where it was found
that values of the peak density lower than about $3\rho_0$
correspond to the liquid state.  We, therefore, take the
value $3 \rho_0$ of the local peak density as 
separating solid- and liquid-like
behaviors.

\begin{figure}
\includegraphics [scale=0.45]{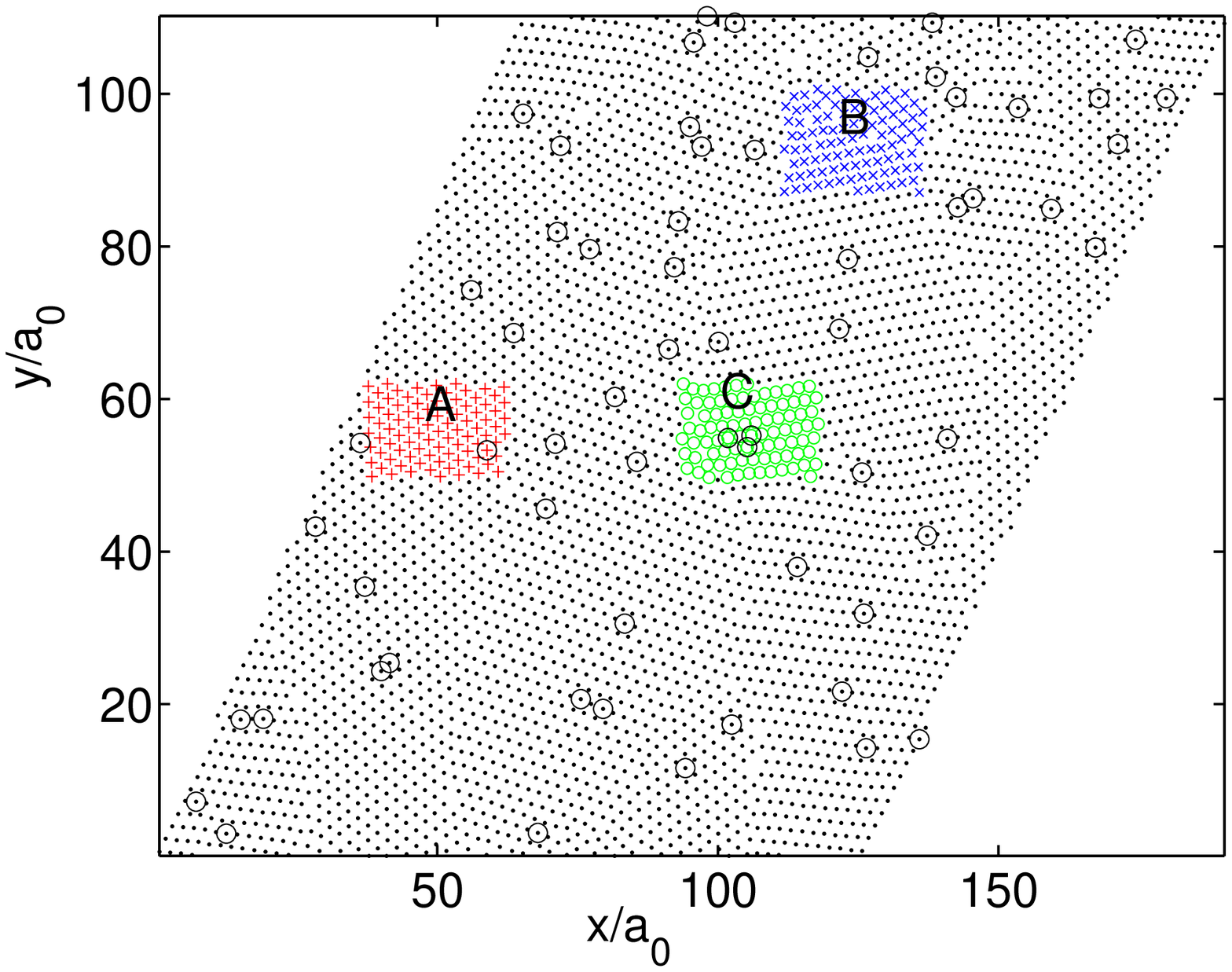}
\includegraphics [scale=0.45]{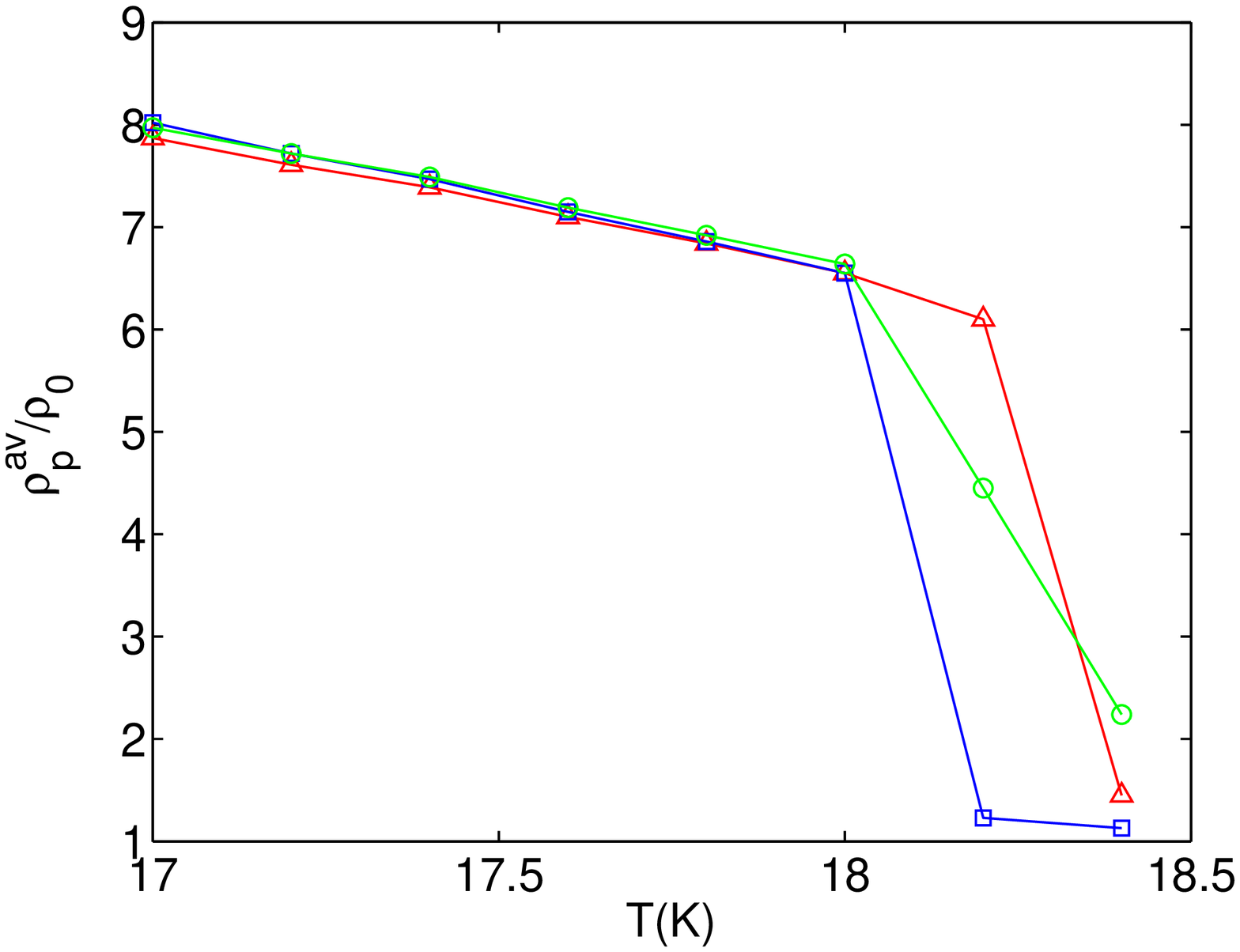}
\caption{\label{fig11} (Color online). Determination of the local melting
temperatures in different regions of a sample. The top panel, at
$T=18.0$K and for the BoG minimum, shows the
three different regions (A, B and C) considered in the calculation. The 
positions of local density peaks in these
regions are indicated by (red) plus signs
(region A), (blue) crosses (region B) and (green) circles (region C). 
The black dots represent the positions of the local density peaks in the
other regions, and the (black) circles show the locations
of the pinning centers. The bottom panel shows the dependence of the
average local peak density $\rho_{av}^p$ (normalized by the liquid density
$\rho_0$) calculated for these regions on the temperature $T$. Data for
regions A, B and C are respectively shown by (red) triangles, (blue) squares,
and (green) circles. The solid lines are guides to the eye.
}
\end{figure}

Using this criterion, we can determine whether a small region of the sample
at a given minimum is in a locally
``solid'' or ``liquid'' state. To do this, we define
a quantity $\rho_{av}^p$ as the average of the local peak densities
in a small region containing $\sim 100$ vortices
(see Fig.~\ref{fig11}). The very high local
density peaks representing vortices trapped at pinning centers are not
included in this average. Values of $\rho_{av}^p$ substantially larger than
$3 \rho_0$ indicate solid-like behavior in the region under consideration,
while substantially lower values of $\rho_{av}^p$ suggest a locally melted
region. To determine how the local melting temperature varies from one part
of the sample to another, we have studied the temperature dependence of
$\rho_{av}^p$  for different regions of the sample. Typical results
are shown in Fig.~\ref{fig11} for the pin configuration for which we have
earlier shown detailed results in several Figures. In the top
panel, where the positions of the local density peaks in the BoG minimum at
$T=18.0$K and the locations of the pinning centers are shown, we
have indicated  three regions for which
the average local peak density $\rho_{av}^p$ was calculated.
The local density peaks in these 
regions are indicated by (red) plus signs
(region A), (blue) crosses (region B) and (green) circles (region C), while
the black dots represent the positions of the local density peaks in the
other regions. The bottom panel of Fig.~\ref{fig11}
shows the dependence of the
$\rho_{av}^p$'s calculated for these three regions on the temperature $T$.
At each value of $T$, the minimum with the lowest free energy
at that temperature was used in the calculation of $\rho_{av}^p$. 

\begin{figure}
\includegraphics [scale=0.45]{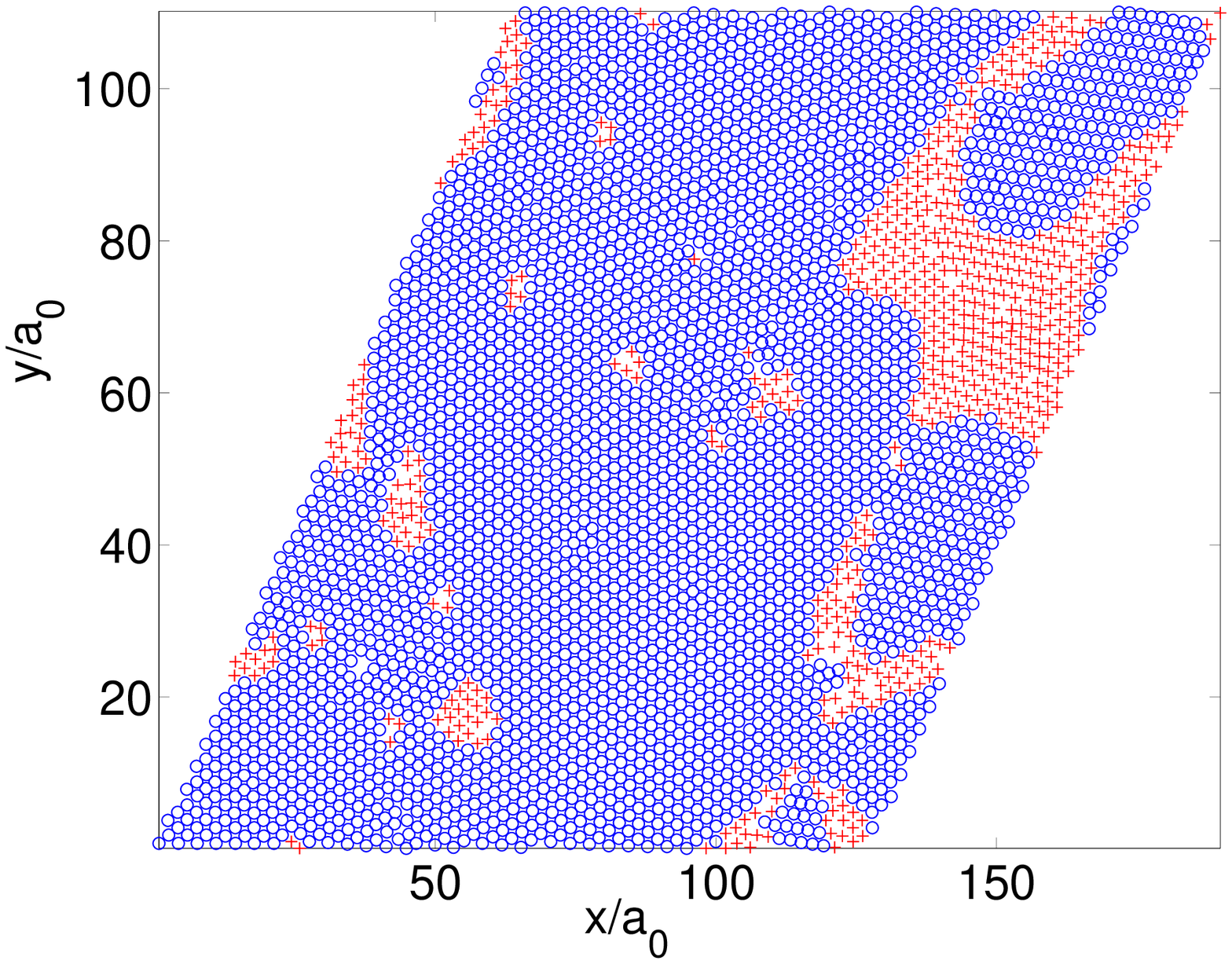}
\includegraphics [scale=0.45]{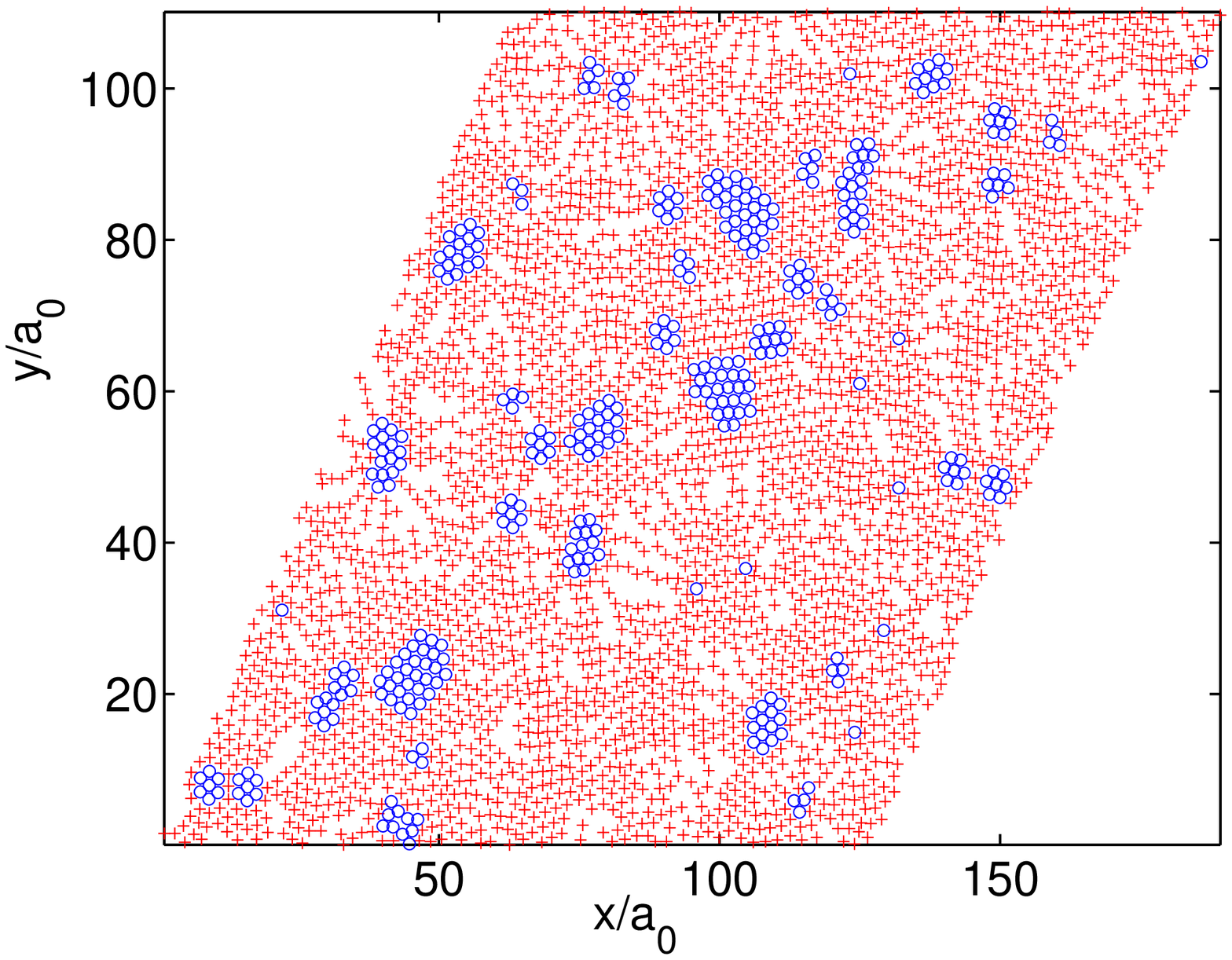}
\caption{\label{fig12} (Color online). Percolation
and melting (see text). The (red) plus signs
denote the locations of liquid-like local density peaks (peak height
smaller than $3\rho_0$), and the (blue) circles indicate the solid-like ones
(peak height greater than $3\rho_0$). The top panel is for the BoG minimum 
of a 4096-vortex sample at 18.2K and the bottom panel is for the IL 
minimum of the same sample at 18.4K. The thermodynamic transition 
between the BoG and IL phases takes place between these two temperatures. 
}
\end{figure}

It is clear from the plot and from similar ones for other
pin configurations (one of which  was shown in 
Ref.~\onlinecite{dv3}) 
that the ``local
melting temperature'', operationally defined as the temperature at which 
$\rho_{av}^p$ drops from
values well above $3 \rho_0$ to values
clearly below, varies from region to region.
The range over which the local melting temperature varies is 
comparable to that found in the experiments~\cite{ban,soi}.
The values of the local melting temperature are strongly correlated with the
local pin structure and the resulting defect structure of the vortex solid.
Region B of the sample does not have any pinning center and is located across
a grain boundary of the BoG minimum. As noted in section~\ref{props}, regions
near grain boundaries appear to melt at temperatures that are slightly lower
than the global transition temperature determined by the crossing of free
energies. This is why the local melting temperature for region B is the lowest.
This temperature corresponds to the transition from the BrG to the BoG state.
In region A  the vortices form a
nearly perfect crystalline arrangement with no topological defects. The local
melting temperature measured in this region is, as a consequence, 
close to that of
the pure system. Region C contains a cluster of 
pinning centers, and the relatively
large value of $\rho_{av}^p$ for this region at temperatures higher than the
BoG to IL transition temperature reflects the local solid-like structure of
vortices situated near pinning centers. Thus, the spatial variation of the
local transition temperature is closely correlated with the ``pinning
landscape'' of the sample.  
  
We have found a second way of correlating the melting transition
with the local density structure of the free energy
minima. 
This second way is based on a
previous study\cite{dv2} of the melting of the vortex
lattice in the presence of periodic pinning, where it
was found that the melting transition 
corresponds to the onset
of percolation of liquid-like regions defined using the 
peak-density criterion mentioned above. 
We focus for this purpose on
the upper transition between
the BoG and IL phases.  
We classify vortices represented by local peaks of the density as solid-like 
or liquid-like, depending on whether the density at the local peak is
higher or lower than $3\rho_0$. We then check whether the regions containing
liquid-like vortices percolate across the sample. We find,
as in the earlier study, 
that the melting transition coincides with the occurrence of percolation
of the liquid-like regions. 
Typical results are shown in Fig.~\ref{fig12}
where the top panel shown the locations of solid- and liquid-like local
density peaks at a BoG minimum obtained at 18.2K (this minimum has the lowest
free energy at this temperature), and the bottom panel shows a similar plot
for the IL minimum of the same sample at 18.4K (the IL minimum is the one with
the lowest free energy at this temperature, so that the melting transition in
this sample occurs between 18.2K and 18.4K). The liquid-like
regions do not percolate at 18.2K, whereas they do at the higher temperature of
18.4K. Thus, the melting transition in this sample also corresponds to the
occurrence of percolation for the liquid-like regions. Similar behavior was
found in the other samples we have studied. This observation provides a 
convenient way of approximately locating the transition point. This may be
useful in other situations where the method of locating the transition
temperature from a crossing of free energies may not be easily implementable
(e.g. at higher pin concentrations where the melting transition is expected 
to become continuous). 

\section{Conclusions\label{conc}}

We have presented here the results of a detailed investigation of 
the structural and thermodynamic properties of a system of vortices in a
highly anisotropic layered superconductor with a small concentration of
randomly placed columnar pinning centers. Both the external magnetic field 
and the columnar pins are assumed to be perpendicular to the superconducting
layers. Our method, based on numerical minimization of the appropriate 
free-energy functional, allows us 
to obtain very detailed information about the density distribution in
the different free energy minima (which allows us to reliably identify
the phases corresponding to these minima), and
to map out the phase diagram of the system. 

There are several salient results of our study. The
first one is the occurrence of a topologically ordered BrG phase at 
low temperatures. While the occurrence of a low-temperature BrG phase in
superconductors with a low concentration of random point pinning centers is
well-established now, relatively little is known about the existence of such
a phase in superconductors with random columnar pins. Our results are
consistent with those of a recent numerical study~\cite{non} of a similar
system. It is clear from
our work that the BrG minima represent a phase  distinct
from the polycrystalline BoG phase also found in our study. 
We
can not conclusively rule out the possibility that free dislocations would 
appear at the nearly crystalline minima
at length scales much longer than those considered in our numerical
study.
If this should happen, then the ``hexatic glass'' phase suggested in 
some earlier
theoretical studies~\cite{chud} would become a possible candidate for
describing the BrG minima found in our study.
This phase, however, would be distinct from the BoG.

Our second important result  is the occurrence of a 
two-step melting transition: we find that 
the low-temperature BrG phase transforms into a polycrystalline BoG phase
as the temperature is increased, and this BoG phase then melts into the
high-temperature IL at a slightly higher temperature.
This conclusion about the occurrence of
two distinct transitions  would remain valid even
if the true nature of the low-temperature phase turns out to be slightly
different from a Bragg glass: our work shows that the BrG 
and BoG minima are quite distinct from each other. The possibility of 
occurrence of a 
two-step melting transition of the vortex lattice in systems 
with random point pinning has been suggested earlier~\cite{men}. Our work
provides support to this suggestion.

To our knowledge, the second
 (lower $T$)
transition between the BoG and BrG phases has not been observed
in experiments on layered
superconductors with a small concentration of random columnar pins.
This may be due to
strong metastability: the BoG minimum 
into which the IL is expected to
freeze as the temperature is decreased
remains locally stable at temperatures
lower than that at which its free energy crosses that of the BrG minimum,
suggesting that it would be difficult to see in experiments 
the transition to the globally stable BrG phase. 
The situation
here may be similar to that found in a recent experimental study\cite{shobo}
of a low-$T_c$ superconductor with weak point pinning which is expected to
exhibit a BrG phase at low $T$. It is, however, found in the experiment that
the vortex solid obtained by cooling the sample in the presence of the
external magnetic field has a polycrystalline structure, indicating that the
metastability of this disordered state prevents the system from reaching the
more ordered (BrG) equilibrium state at low temperatures. 
In our numerical work, the
BrG minima were obtained by performing the free-energy minimization from
an initial configuration with crystalline order. It is not clear
how a similar procedure can be adopted in experimental studies. 

A low-temperature BrG phase has been observed in a
recent simulation\cite{non} of a model of HTSC with a small concentration of
columnar pinning centers. However, this simulation finds a 
single first-order transition between the BrG and IL phases at low pin
concentrations: the small intermediate region of 
BoG phase found in our study is not observed. This is probably
due to the smallness of the system sizes ($\sim$ 100 vortex lines) used in 
the simulation. As discussed in Sec.~\ref{phasediag} above, the polycrystalline
BoG minima are found only if the sample size is larger than the typical size of
the crystalline domains: only the BrG minimum is found at low temperatures if
this condition is not satisfied. 
Since the crystalline domains become large at small
pin concentrations, it is quite likely that a simulation with small system
sizes and low pin concentration would not see the BoG phase. Another
possibility is that a narrow ``two-phase'' region found 
in Ref.~\onlinecite{non} near the BrG melting transition actually corresponds
to the intermediate BoG phase found in our study. A third possibility is that
the intermediate BoG phase, which arises due to a competition between the
elastic and pinning components of the free energy (see Sec.~\ref{phasediag}
where it is shown that the pinning energy favors the BoG phase),
does not appear in the simulation of  Ref.~\onlinecite{non}
because the pins are assumed to be ''weak'' in that work.

Our study also illustrates the spatial inhomogeneity of the melting process
in the presence of disorder: we have found that a ``local'' melting 
temperature defined using a criterion based on the degree of localization of
the vortices shows considerable spatial variation. This spatial variation
is strongly correlated with the ''pinning landscape'' associated with the
random spatial location of the pinning centers. 
Our results about the spatial inhomogeneity
of the local melting temperature 
are consistent with those of recent experiments
\cite{ban,soi} on superconductors with pinning disorder.

These results establish the usefulness of our numerical method in dealing with
the problem of vortex matter in the presence of random pinning. 
It would be interesting to examine the extent to which our results 
depend on the values
of the parameters appearing in the free energy functional.
It would also be
clearly useful to carry out similar studies of other related problems, such
as the complete phase diagram of systems with random columnar 
pins in the $T-c$ plane, and the phase behavior of systems with random
point pinning. Some of these investigations are currently in progress.

\begin{acknowledgments}
We thank G. I. Menon for several useful discussions.
\end{acknowledgments}


\end{document}